\documentclass[twocolumn,english,prx,showpacs,longbibliography]{revtex4-1}
\usepackage{amssymb}
\usepackage{graphicx}
\usepackage{dcolumn}
\usepackage{bm}
\usepackage{amsmath}

\def\OMIT#1 {{}}
\def\MEMO#1 {{}}

\newcommand{\sjtu} {Key Laboratory of Artificial Structures and Quantum
Control, Department of Physics and Astronomy, Shanghai Jiao Tong University, Shanghai 200240, People's Republic of China}

\newcommand{\iop} {Beijing National Lab for Condensed Matter Physics and Institute of Physics,
Chinese Academy of Sciences, Beijing 100190, China}

\begin{document}
\title{Approximating quantum many-body wave-functions using artificial neural networks }

\author{Zi Cai}
\email{zcai@sjtu.edu.cn}
\affiliation{\sjtu}

\author{Jinguo Liu}
\affiliation{\iop}

\date{ \today }

\begin{abstract}
In this paper, we demonstrate the expressibility of artificial neural networks (ANNs) in  quantum many-body physics by showing that a feed-forward neural network with a small number of hidden layers can be trained to approximate with high precision the ground states of some notable quantum many-body systems. We consider the one-dimensional free bosons and fermions, spinless fermions on a square lattice away from half-filling, as well as  frustrated quantum magnetism with a rapidly oscillating ground-state characteristic function.  In the latter case, an ANN with a standard architecture fails, while that with a slightly modified one successfully learns the frustration-induced complex sign rule in the ground state and approximates the ground states with high precisions. As an example of practical use of our method, we also perform the variational method to explore the ground state of an anti-ferromagnetic $J_1-J_2$ Heisenberg model.
\end{abstract}

\maketitle

\section {Introduction}
A central challenge in quantum many-body physics is developing efficient numerical tools for strongly correlated systems, whose Hilbert space dimensionality grows exponentially with the system size, so does the information required for characterizing a generic state of the system. However, for many physical systems of practical interest, the ground states may have a simplified structure, thus, it can be appropriately approximated using an exponentially smaller number of parameters than that required for characterizing generic states. Typical examples include low dimensional strongly correlated systems, whose ground states can be represented in terms of matrix product states by taking  advantage of limited entanglement entropy in the ground states\cite{White1992,Verstraete2008,Schollwock2011}. For higher-dimensional systems, a different routine involving stochastic sampling applies for  certain types of strongly correlated systems with positive-definite ground states, with Quantum Monte Carlo (QMC) algorithm providing a well-controlled method to evaluate the physical quantities of interest based on an exponentially small fraction of all possible configurations\cite{Sandvik1991,Prokofev1998,Gubernatis2016}. In spite of the remarkable achievement of these numerical methods, developing a general strategy to represent typical many-body wave-functions that bypasses the exponential complexity remains a formidable, if not impossible, task and is a question of principal interest in the condensed matter community.

In general, a many-body wave-function can be expanded in terms of a set of orthogonal basis (e.g. the Fock basis) and can be fully characterized by a function where one feeds in a basis and gets the output the set of corresponding coefficient in the wave function.  Consider a spin lattice system as an example; in this case,  an arbitrary wave-function can be expanded as: $\Psi=\sum_{\boldsymbol{\sigma}}C[\boldsymbol{\sigma}]|\boldsymbol{\sigma}\rangle$,  where $|\boldsymbol{\sigma}\rangle=|\sigma_1\dots\sigma_L\rangle$ are $\{S^z_i\}$-eigenstates spanning the Hilbert space of the spin configurations. Therefore, the task amounts to efficiently approximating the function $C[\boldsymbol{\sigma}]$ using an exponentially smaller number of parameters than the Hilbert space dimensionality$\sim 2^{L}$. Among the existing modern techniques used for approximating functions, artificial neural networks (ANNs), as a powerful tool for data fitting and feature extraction, have not only achieved remarkable successes in machine learning and cognitive science fields in the past decades\cite{Nielsen2015,Goodfellow2016}, but have also recently attracted considerable attention of researchers in condensed matter community. Applying machine learning methods to problems in condensed matter physics is not only interesting on its own right\cite{Mehta2014,Lin2016,Chen2017,Stoudenmire2016}, but it may also potentially provide new ideas and have practical applications for solving  complex physics problems, such as identifying classical and quantum phases of matter and locating the phase transition points\cite{Carrasquilla2017,Wang2016,Nieuwenburg2017,Deng2016b,Torlai2017,Ohtsuki2016,Zhang2016,Ch2016,Ponte2017}, categorizing and designing
materials\cite{Curtarolo2003,Kalinin2015,Ghiringhelli2015},  improving existing numerical techniques\cite{Broecker2016,Huang2016,Huang2017,Liu2016,Liu2017} and even developing new methods in the quantum many-body physics\cite{Carleo2017,Wang2017}. Owning to its tremendous capability in function approximations, ANNs can  also be considered as  novel representations of  many-body wave-functions\cite{Carleo2017,Gao2017,Deng2016,Huang2017b,Levine2017}, e.g. in a seminal work, Carleo {\it et al} proposed  a new kind of variational wave functions using the restricted Boltzmann machine\cite{Carleo2017}.


Even though it can be prove mathematically that ANNs can in principle approximate any smooth function to any accuracy\cite{Kolmogorov1961,Hornik1989,Cybenko1989}, what matters in practice is the efficiency of the method: the amount of resources an ANN needs to approximate a given multi-variable function.    In this paper, we will demonstrate the expressibility of neural networks in approximating and characterizing  quantum many-body wave-functions using some notable examples of physical interest.    For a wave function with completely random coefficients, the information encoded in it cannot be compressed; thus, an exponentially large number of parameters are needed.  The goal of this paper is to express the ground-states of several notable many-body systems in terms of neural networks of feasible size and a small number of hidden layers, and, most importantly, networks learns in a reasonable time. The success of this approach relies on the specialty of the ground state compared to a generic eigenstate, where the underlying physical laws encoded in the ground-state's wave function can be extracted by the neutral networks through big amount of training.
\begin{figure}[htb]
\includegraphics[width=0.9\linewidth,bb=1 6 360 206]{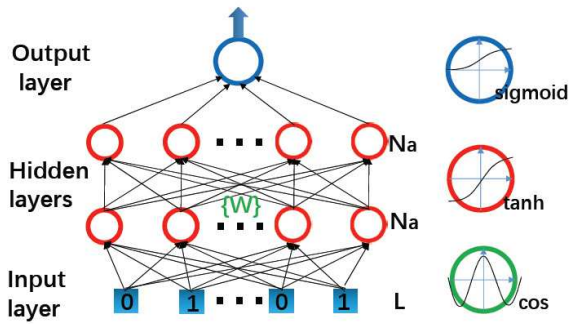}
\caption{ (Left) The structure of the feed-forward ANN we used to approximate  quantum many-body groundstate wavefunctions; (Right) three types of neurons with different activation functions.}
\label{fig:fig1}
\end{figure}

In the following, we show which ground-state wave-functions can be efficiently expressed by a simple neural network, and for those who can't, how the ANN's architecture should be modified to achieve this goal. The scaling of the computation resources with the system size is also investigated.  Special attention is devoted to  wave functions with the \textquotedblleft sign problem\textquotedblright, where the function $C[\boldsymbol{\sigma}]$ may alter its sign even for a local change in the input $[\boldsymbol{\sigma}]$, and thus cannot be considered  a \textquotedblleft smooth\textquotedblright function. The sign problem not only hinders the application of QMC methods, but also make it more difficult for  a simple ANN to learn a wave-function. However, as we shown, this problem can be circumvented by dividing the task into two sub-tasks and by designing  ANNs with corresponding architectures. This is a typical example that illustrates the neural network's ability of extracting the physics laws, even for those too complex to be written explicitly.

The rest of the paper is organized as follows:  first we introduce the structure of the ANN then used it to investigate some notable examples, including  one- (1D) and two-dimensional(2D) free bosons and fermions, whose exact ground states are compared to those predicted by the ANN. To determine whether the ANN approach can work for large systems, we adopt the importance sampling algorithm to calculate physical quantities instead of wave functions, and compare them to the exact ones. Then, we attack the most difficult part: approximating the ground state of a frustrated quantum magnetism whose characteristic function can dramatically change its sign, thus being very different from a smooth function, which makes it extremely difficult for a regular ANN to approximate. In spite of this, we find that a slight modification of neurons in the ANN allows to capture the sign rule of the frustrated quantum magnetism, even at the phase transition point. Finally, we  discuss the practical application of this method based on the variational method.

\section {Methods}
Before we proceed further to discuss specific examples, let us describe the details of the ANNs and the optimization methods we will use. We consider a fully connected feed-forward neural network  consisting of interconnected group of nodes (neurons) with a stacked layered structure, and its expressibility is encoded in sets of adaptive weights of  connections between neurons in adjacent layers. As shown in Fig.\ref{fig:fig1}, we consider a four-layer ANN, with two hidden layers (each containing $N_b$ neurons) that are sandwiched between the input layer (accepting the Fock basis ($[\boldsymbol{n}]$ or $[\boldsymbol{\sigma}]$) and the output layer that output the corresponding coefficient predicted by the ANN ($C_P[\boldsymbol{\sigma}]$) .  A neuron can be considered to be an elemental processor:
\begin{equation}
o^{[n+1]}_i=f(\sum_j W^{[n]}_{ij}o^{[n]}_j+b^{[n]}_i),
\end{equation}
 where $o^{[n]}_j$ is the output of the $j$-th neuron of the $n$-th hidden layer, $W^{[n]}_{ij}$ denotes the connection weight between the $n$-th and $n-1$-th hidden layers, and $b^{[n]}_j$ is the bias in this neuron.    The activation function $f(x)$ can be  any smooth nonlinear function. However, as we show below, choosing a proper nonlinear function may significantly increase the ANN's efficiency for approximating certain target functions. We introduce a fidelity function:
\begin{equation}
F=1-|\sum_{\boldsymbol{\sigma}}\langle C_{T}^*[\boldsymbol{\sigma}]C_{P}[\boldsymbol{\sigma}]\rangle|
\end{equation}
to measure the difference between the target function $C_{T}[\boldsymbol{\sigma}]$ and the one predicted by the ANN $C_P[\boldsymbol{\sigma}]$. Here we should point out that to calculate this quantity, the target functions are known in advance.  We will discuss how to generalize the current method  to explore new quantum many-body systems with previously unknown ground states in Sec.\ref{sec:variational}.  The problem now reduces to an optimization problem with the goal of finding the minimum of $F$ in the landscape of ANN parameters (weights of the connection and bias, which are denoted as $\{W\}$ in the following). In the following, we adopt the ANN construction methods and the optimization techniques that are readily available in the machine learning libraries TensorFlow\cite{Abadi2015}, with the training time  measured in the units of $T_0$, corresponding to the period of a single  optimization  iteration  that depends on the details of the ANN, more details about the initialization and training can be found in the Appendix.
\begin{figure*}[htb]
\includegraphics[width=0.325\linewidth,bb=46 234 488 554]{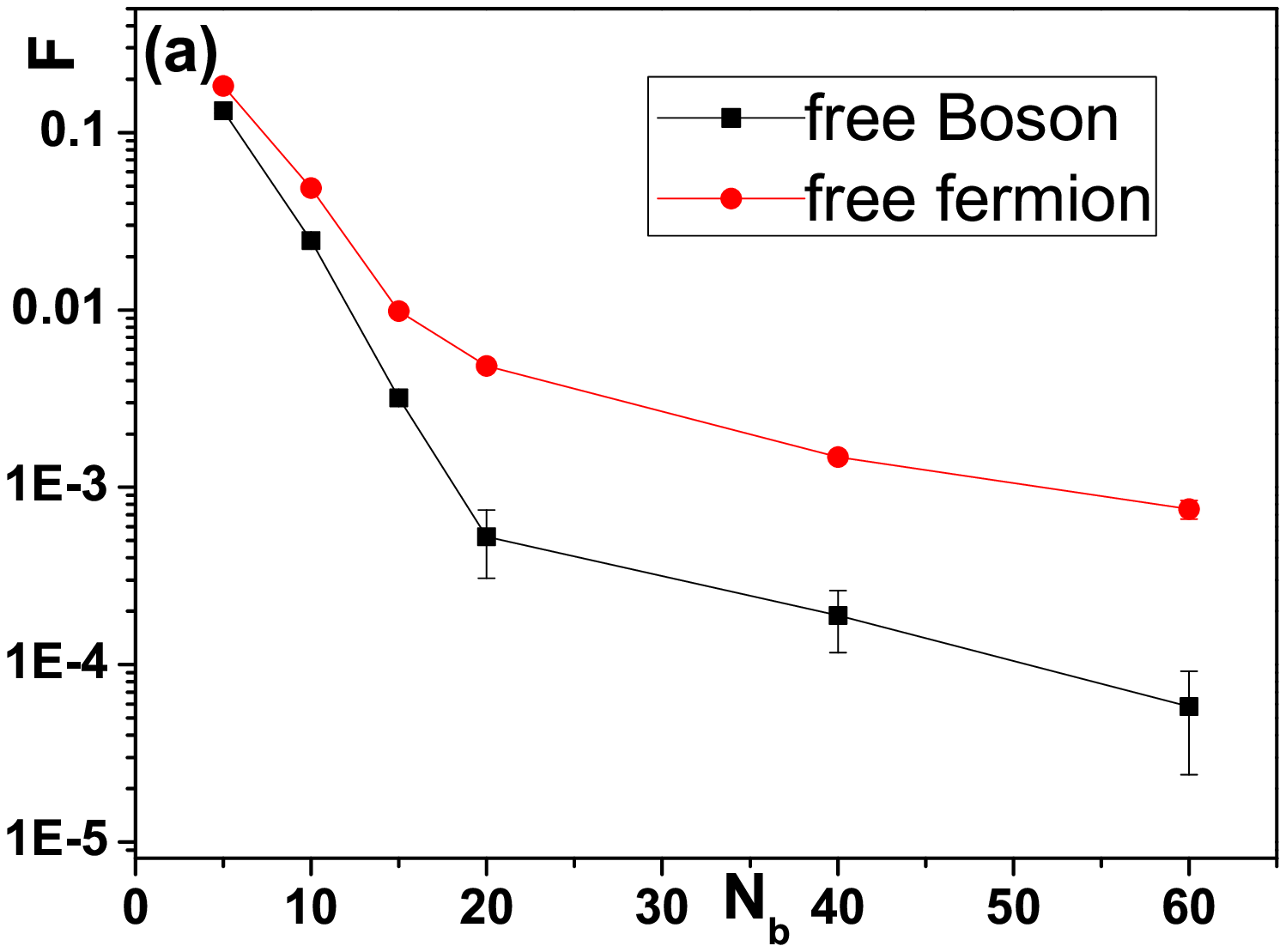}
\includegraphics[width=0.325\linewidth,bb=15 18 278 212]{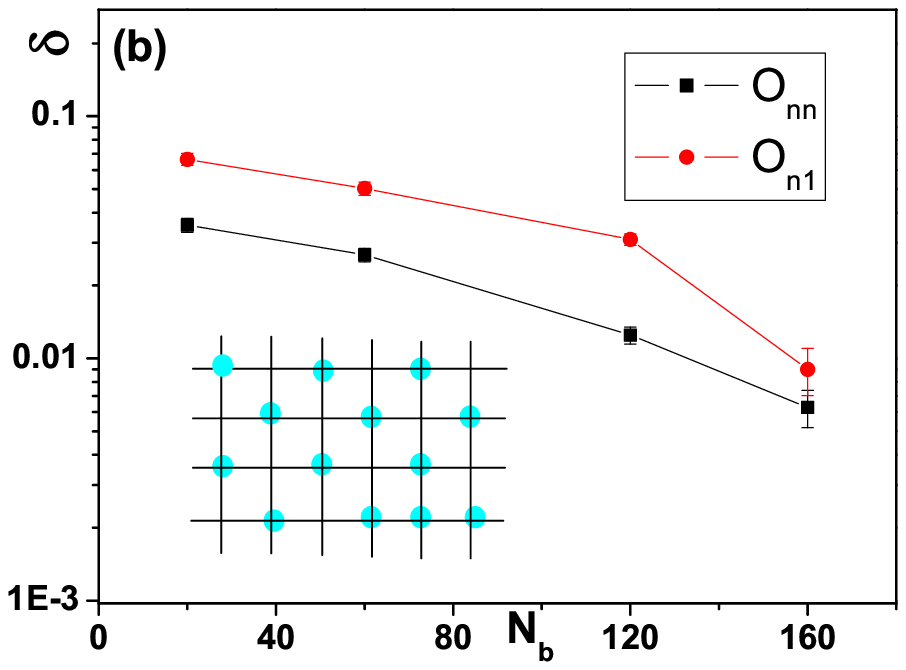}
\includegraphics[width=0.325\linewidth,bb=46 234 488 554]{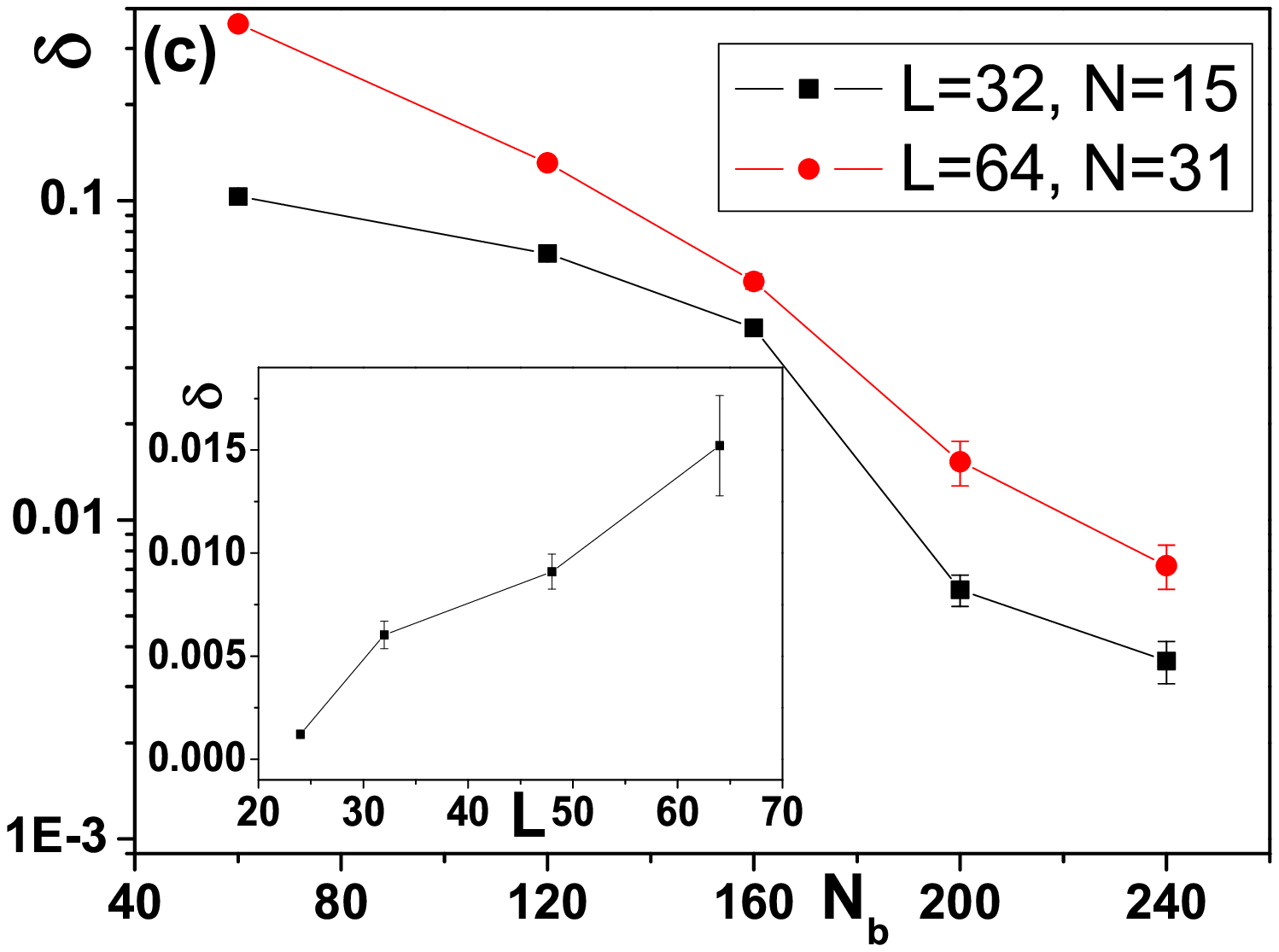}
\caption{(a) Fidelity $F=1-|\langle \Psi_{ED}|\Psi_{ANN}\rangle|$ as a function of the number of neurons $N_b$, for 1D free bosons with $L=12$, $N=12$ and the corresponding Hilbert space dimensionality $\mathbb{D}=1352078$ and for 1D free fermions with $L=24$, $N=11$ and $\mathbb{D}=2496144$;  (b) the precision function $\delta=|O_{ANN}-O_{ED}|/O_{ED}$ for two different quantities $O_{nn}=\frac 1L \sum_i\langle n_i n_{i+1}\rangle$ and $O_{n1}=\langle n_1\rangle$ for  2D free fermions with L=24($4\times6$), N=13 and  $\mathbb{D}=2496144$, the inset shows a typical Fock configuration; (c) the precision function $\delta$ for $O_{nn}=\frac 1L \sum_i\langle n_i n_{i+1}\rangle$ as a function of $N_b$ obtained by the importance sampling algorithm for a 1D free fermion, where both the amount of the training and the sampling data are chosen as $\mathbb{H}=10^6$ in the simulations of various system sizes, the inset show $\delta$ as a function of $L$ with a fixed $\mathbb{H}=10^6$ and $N_b=200$. }
\label{fig:fig2}
\end{figure*}

\section {Free bosons/fermions systems}
\label{sec:free}
The first class of wave-functions that we investigate is the ground states of the simplest many-body systems composed of free bosons or fermions. In spite of their extreme simplicity, these wave-functions are perfect touchstones to test the expressibility of the ANN, for two reasons: first the well-known analytic forms of these wave functions serve not only  as a target function during the training process of a neural network, but also a tester of the accuracy of its predictions; Second, in some cases, {\it e.g.} the ground state of 2D free fermions away from half-filling, the wave function is simple but not trivial, in the sense that it is difficult to be characterized using the existing numerical methods like the MPS and the path integral QMC method, because this wave-function suffers from the entanglement area law and the sign problem simultaneously.

The examples we studied in this section include the ground state of free bosons in 1D lattice, and those of free fermions in 1D and 2D lattices, with the  filling factor away from half-filling. For free bosons (FB) in a 1D lattice with length L and unit filling factor (with the overall number of bosons $N=L$), the ground state wave function can be written as: $\Psi_{FB}=\sum_{\boldsymbol{n}}C_{FB}[\boldsymbol{n}]|\boldsymbol{n}\rangle$  where $|\boldsymbol{n}\rangle=|n_1\dots n_L\rangle$ is the occupation number basis spanning the Hilbert space under the constraint $N=L$. The analytic form of the multi-variable characteristic function is: $C_{FB}[\boldsymbol{n}]=\sqrt{L!/n_1!\dots n_L!}/L^{L/2}$. The wave-function of a free fermion lattice system has a similar form while $n_i$ can only be  0 or 1. We assume there are N particles and $C_{FF}[\boldsymbol{n}]=\det[\mathbb{M}]$, where $\mathbb{M}$ is a $N\times N$ matrix with the matrix elements $M_{ij}=f_i(x_j)$, with $x_j$ denoting the position of the $jth$ fermion, and $f_i(x)$ denoting the $ith$ single-particle eigenstate, in a 1D chain with a periodic boundary condition (PBC), $f_i(x)=\frac{1}{\sqrt{L}}e^{ik_i x}$ with $k_i$ being the $ith$ momentum. In a 2D lattice, both $x_j$ and $k_i$ are replaced by a 2D vector  $\boldsymbol{x_j}$ and $\boldsymbol{k_i}$. Based on the above exact results,  we implemented the training process, aiming to minimize the fidelity function $F$ by adjusting the parameters of the ANN. To avoid the problem of over-fitting,  the number of the variational parameters in the ANN was chosen to be on  the order of $\mathcal{O}(N_b^2)\sim 10^3$, significantly smaller than the typical Hilbert space dimensionality of the systems we studied here($\sim 10^6$).  As shown in Fig.\ref{fig:fig2} (a), for cases of 1D free bosons or fermions, an ANN with only a few neurons $N_b \sim\mathcal{O}(10^1)$ can easily approximate the corresponding target function with an extraordinarily high precision. For 2D fermions, a direct approximation of $C_{FF}[\boldsymbol{n}]$ based on the ANN with the current structure seems to fail because the sign of $C_{FF}[\boldsymbol{n}]$ can dramatically change owning to the fermionic statistics in two dimensions.  We will reconsider this point later on. To avoid this problem, here we chose $|C_{FF}[\boldsymbol{n}]|$ instead of $C_{FF}[\boldsymbol{n}]$  as our target function, which still allowed us to calculate the average values of diagonal operators in the Fock basis  ({\it e.g.} the nearest neighbor (NN) density correlation $O_{nn}=\frac 1L \sum_i\langle n_i n_{i+1}\rangle$ and the local density operator $O_{n1}=\langle n_1\rangle$). As shown in Fig.\ref{fig:fig2} (c), an ANN with $N_b\sim\mathcal{O}(10^2)$ can give rise to values of $O_{nn}$ and $O_{n1}$ with  precisions $\sim \mathcal{O}(10^{-3})$. We also notice that the ground state of interacting quantum models (e.g. the quantum spin model\cite{Carleo2017} and bose-Hubbard model\cite{Saito2017,Saito2018}) have been studied by other machine learning methods, where the ANNs are trained to minimize the ground state energy, instead of the fidelity of the wave function.

In all of the above-studied cases, the system size is relatively small since we want to compare our results with the exact ones. In what follows, we show that in principle, there is no intrinsic difficulty to use ANNs to simulate larger systems, whose Hilbert space dimensionality is much larger than the memory of any computer, thus making it impossible to store the predicted wave functions and calculate its overlap with the exact ones. As a consequence, we focus on physical quantities in the ground states instead of focusing on the wave functions themselves, to test the accuracy of the ANN. The key point is that we approximate the characteristic functions only training the ANN only over a tiny fraction of the entire configuration space and we assume that the predicted functions are also valid for other configurations once the ANN finds the correct form of the target function. Obviously, the efficiency of the function approximating strongly depends on the choice of  training configurations, which need to be \textquotedblleft representative\textquotedblright in the Hilbert space. In other words, the probability of choosing a certain configuration $[\boldsymbol{n}]$ should be proportional to its weight $|C[\boldsymbol{n}]|^2$ in a given wave function, which can be achieved by the importance sampling in the Monte Carlo algorithm. Consider the ground state of a 1D $L=64$, $N=31$ free fermion system as an example. For this system, we first implemented importance sampling to generate  millions of \textquotedblleft representative\textquotedblright configurations based on the exact value of $|C_T[\boldsymbol{n}]|^2$, and use these generated configurations to train the ANN and approximate the target function. After completing the training, a new set of \textquotedblleft representative\textquotedblright configurations was generated according to their weights predicted by the ANN $|C_P[\boldsymbol{n}]|^2$, and we calculated the valued of the physical quantities based on these new configurations and compared them with their exact values. The result is shown in Fig.\ref{fig:fig2} (c), where we observe that the precision of $O_{nn}$ evaluated based on the above-described sampling scheme can reach  $ \mathcal{O}(10^{-3})$. For a fixed amount of training and sampling data, the scaling relation between the precision and system size is shown in the inset of Fig.\ref{fig:fig2} (c).

\begin{figure*}[htb]
\includegraphics[width=0.32\linewidth,bb=12 19 277 212]{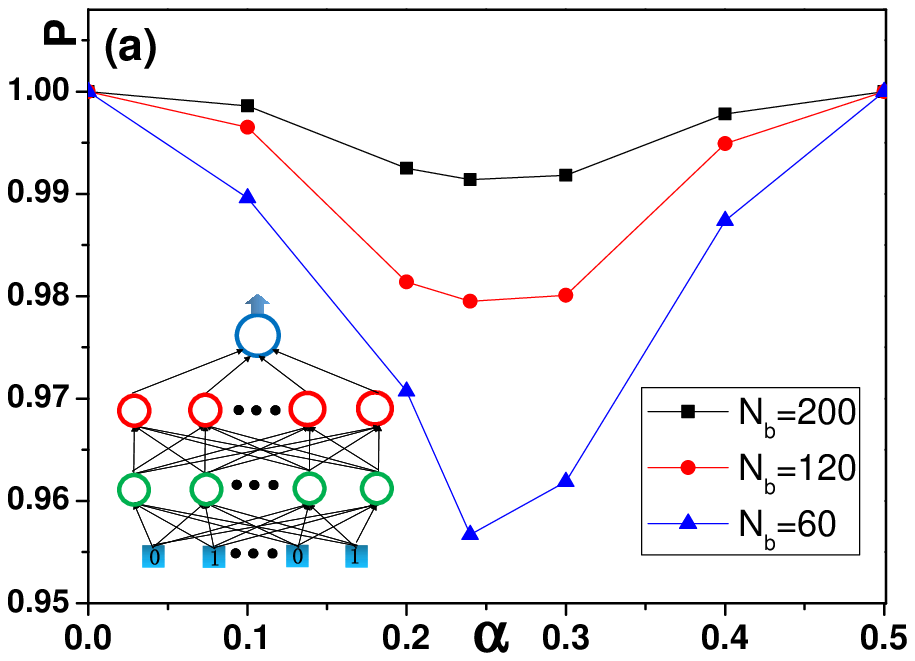}
\includegraphics[width=0.32\linewidth,bb=60 230 485 560]{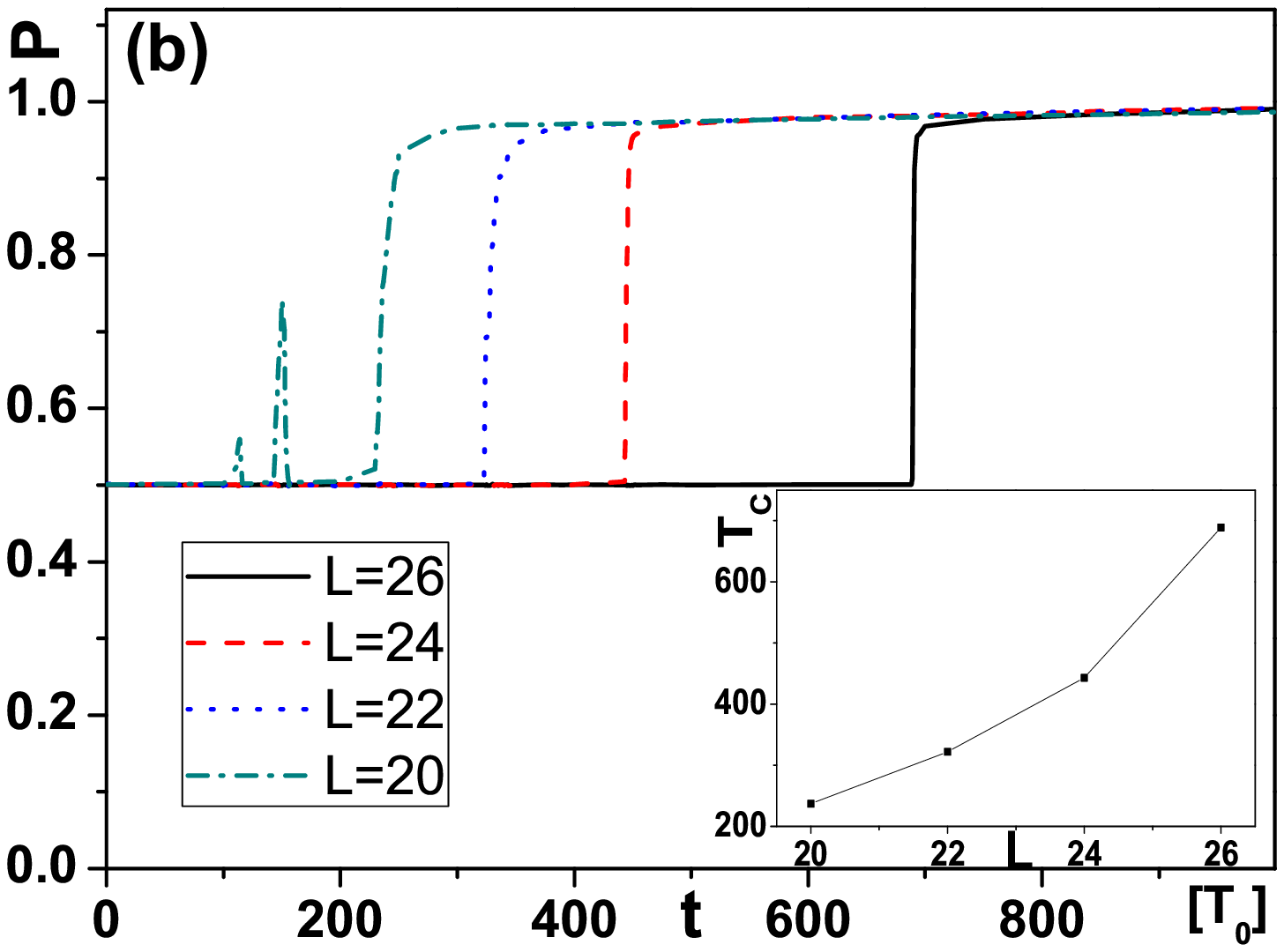}
\includegraphics[width=0.32\linewidth,bb=54 228 487 544]{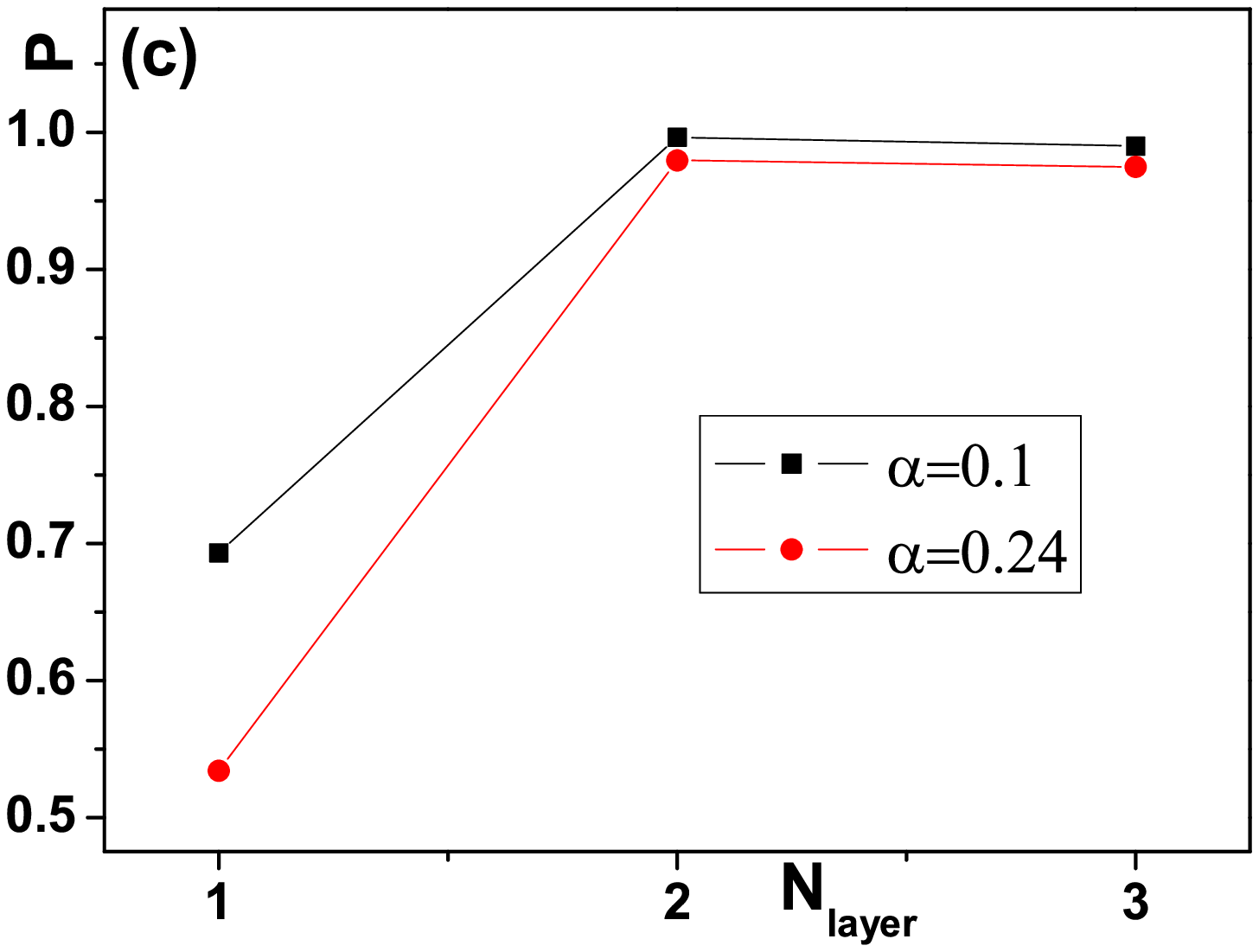}
\includegraphics[width=0.32\linewidth,bb=44 229 488 545]{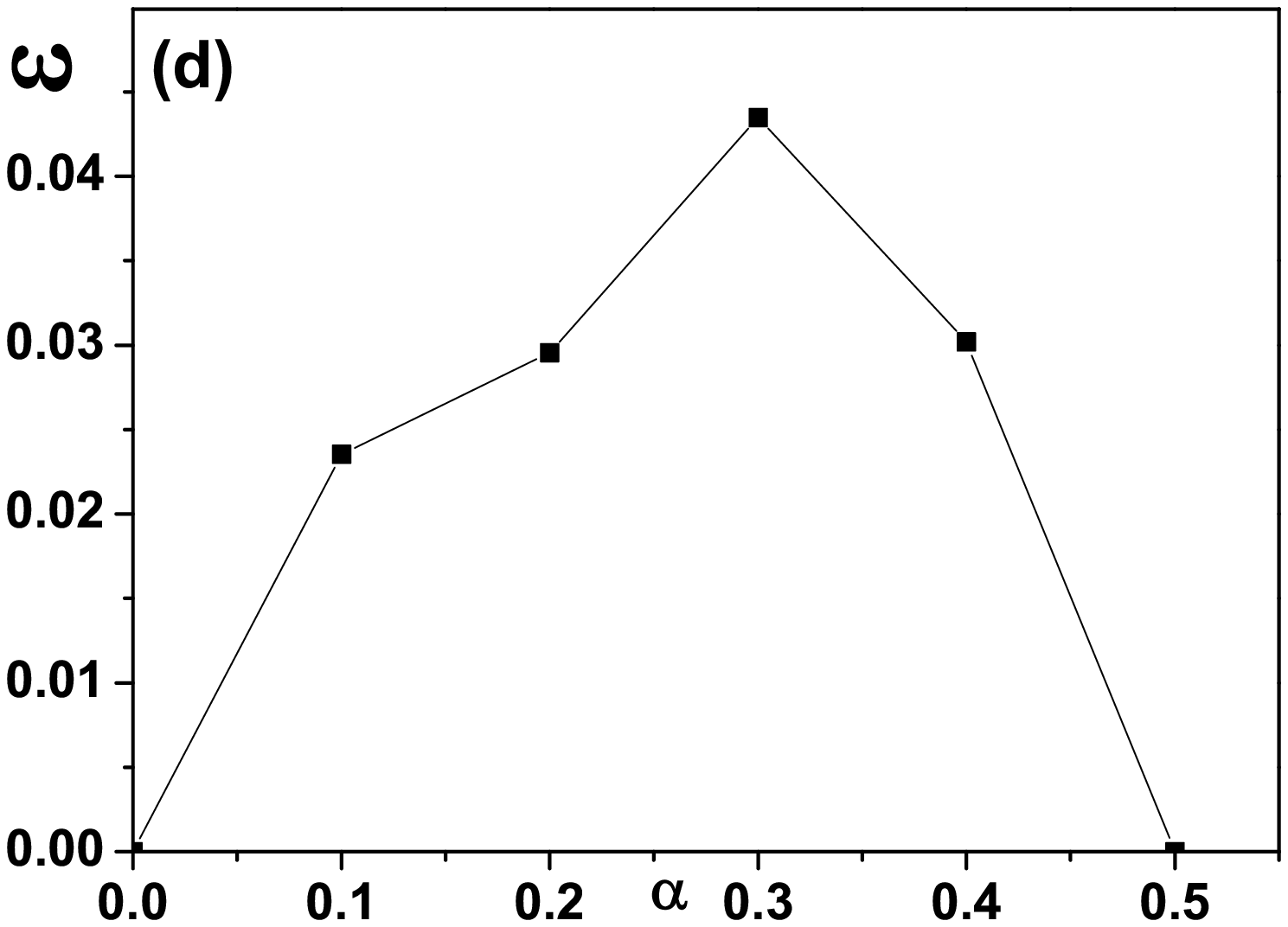}
\includegraphics[width=0.32\linewidth,bb=44 230 485 545]{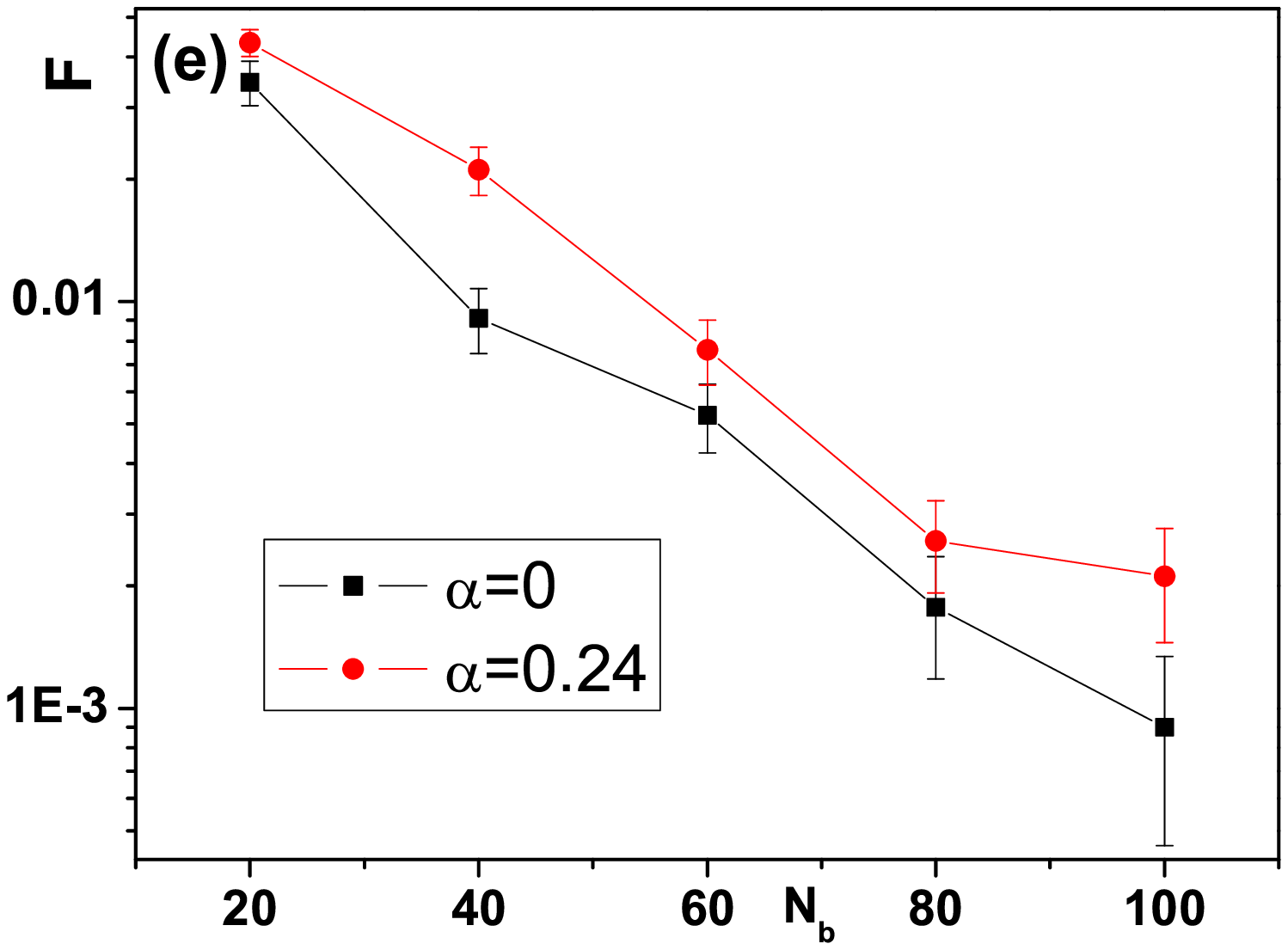}
\includegraphics[width=0.32\linewidth,bb=54 228 487 544]{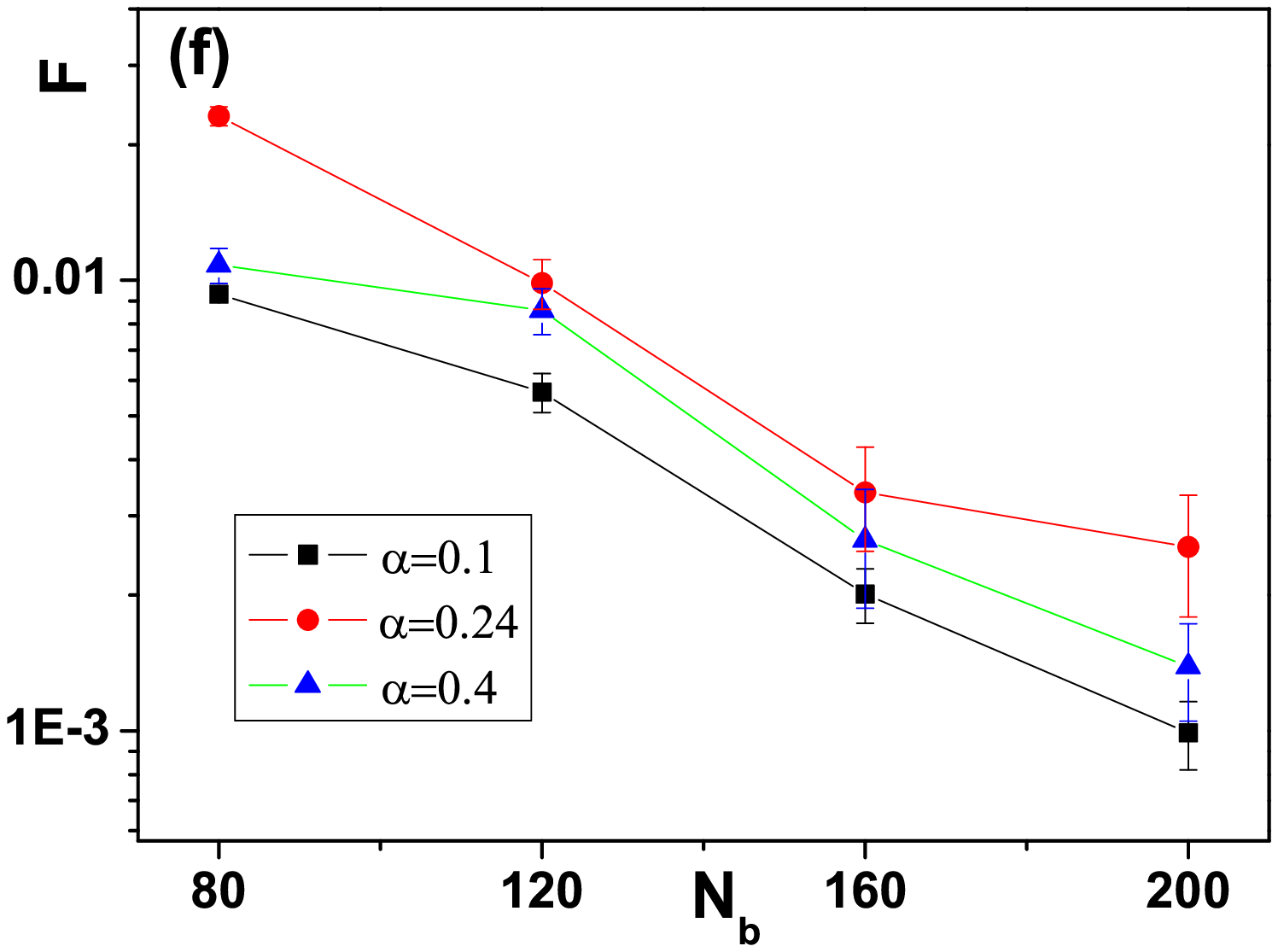}
\caption{(a)  The accuracy function $P(\alpha)$ of the sign rule  in the ground state of 1D $J_1-J_2$ model with $L=24$ and various $N_b$, with the inset showing the modified ANN used to capture the sign rule with  neurons in the first hidden layer replaced by those with the cosine activation functions $f(x)=\cos(\pi x)$; (b) The evolution of P during the training with fixed $N_b=200$ and various systems sizes, with the inset showing a typical \textquotedblleft confusion\textquotedblright  time $T_c$ in the training as a function of system size with $N_b=200$; (c) The accuracy $P$ predicted by the sign-ANN as a function of the layer of the ANN with a fixed $N_b=120$; (d)the ratio between the average absolute value of the coefficient with the erroneously assigned sign and that of the whole training set with $N_b=120$; (e) The fidelity as a function of $N_b$ for the amplitude ANN.    (f) The Fidelity in the $J_1-J_2$ model and various $\alpha$  as a function of $N_b$  predicted by the combination of the sign (with $N_b^s=N_b$) and amplitude ANNs (with $N_b^a=N_b/2$).  For (c)-(f), the system size $L=24$. }
\label{fig:fig3}
\end{figure*}

\section {Non positive definite wave functions}
\label{sec:Frustration}
In general, neural networks perform much better on approximating  smooth functions  rather than  rough ones, which may restrict their applicability in quantum many-body physics, because for certain types of important wave-functions, their characteristic functions may drastically change and even alter its sign in response to a slight change in the input configuration. The sign problem, in its different forms, imposes challenges on both existing methods ({\it e.g.} QMC) and the current neural network function approximating. For example, the ANN with a simple structure illustrated in Fig.\ref{fig:fig1} fails to approximate the ground state of one of the simplest models: a 1D anti-ferromagnetic Heisenberg model. In QMC, the sign problem in such a bipartite lattice can be avoided by performing a basis rotation. In the current method, we attack the problem by separately approximating the amplitude and the sign of the target function by two different ANNs. The ANN that approximates the amplitude is similar to those studied above, with the only difference that the target function is replaced by its absolute value. This \textquotedblleft amplitude ANN\textquotedblright performs as well as that in the previously described ones in Sec.\ref{sec:free}.

The difficult part is to approximate the sign function: $S[\boldsymbol{\sigma}]=(C[\boldsymbol{\sigma}]/|C[\boldsymbol{\sigma}]|+1)/2$, which takes the values of 1(0) if $C[\boldsymbol{\sigma}]$ is positive (negative). For a 1D lattice (or more generally, a bipartite lattice) model, it is well-known that the sign of $C[\boldsymbol{\sigma}]$ obeys the Marshall sign rule: $S[\boldsymbol{\sigma}]=1/0$ if in $\boldsymbol{\sigma}$ the total number of down spins in the odd sites is even/odd. This mathematical theorem enables us to perform a basis rotation in the even or odd sites to eliminate the sign problem in QMC simulations. For the current method, the question is without the prior knowledge of the  Marshall sign rule, whether a ANN can automatically extract it from the training set data? Based on our numerical tests, we found that a standard ANN, as the one shown in Fig.\ref{fig:fig1} fails to extract the Marshall sign rule, but a modified ANN with the activation function of neurons in the first hidden layer replaced by a cosine function (hereafter denoted as \textquotedblleft sign ANN\textquotedblright) succeeds. This modification can significantly increase the efficiency of the ANN with the accuracy of $100\%$  because the  cosine function is more capable of capturing the even/odd features in the input data. Another important wave-function is the ground state of the Majumdar-Ghosh model:  the dimerized state: $\Psi=\bigotimes_{i=1}^{\frac L2} \frac 1{\sqrt{2}}[|\uparrow\rangle_{2i-1}|\downarrow\rangle_{2i}-|\downarrow\rangle_{2i-1}|\uparrow\rangle_{2i}]$, with the sign function $S[\boldsymbol{\sigma}]=[\prod_{i=1}^{\frac L2} (S^z_{2i-1}-S^z_{2i})+1]/2$. We found that this sign rule can also been satisfactorily learned by the sign ANN with the accuracy of $100\%$.

In the two examples above, the sign rules are rather simple in the sense that they can be written explicitly in a simple form. However, for a generic wave-function with the sign problem, this is not the case. The sign rule may be too complex to be captured by  programming or designing explicit algorithms, which on the other hand is exactly what ANNs are good at. Approximating the sign function becomes a classification problem, which reminds us of one of the most successful applications of ANN: recognizing handwritten digits.  In this classical problem,  the ANN was shown to automatically and successfully infer the rules of classification using the training set examples. Here we adopt a similar strategy to extract the elusive sign rule for the ground state of a frustrated quantum magnetism model: the 1D $J_1-J_2$ AF-Heisenberg model with the Hamiltonian: $H=\sum_i[J_1 \boldsymbol{S}_i\boldsymbol{S}_{i+1}+J_2 \boldsymbol{S}_i\boldsymbol{S}_{i+2}]$  (where both $J_1$ and $J_2$ are non-negative). The two cases studied above are exactly the ground state of this model in two limits: $\alpha=0$ and $0.5$ with $\alpha=J_2/J_1$ .

For a general $\alpha$, there is no exact solution of the ground state; thus, we used the Lanczos method to calculate the sign function for a finite size system and compare it to the one predicted by the sign ANN.   The accuracy {\it v.s.} $\alpha$  is plotted Fig. \ref{fig:fig3} (a) for various $N_b$, and we observe that in the entire region $0\leq \alpha\leq 0.5$,  the accuracy predicted by the sign ANN is high, reaching $99\%$, while the minimum of the accuracy corresponds to the phase transition point. Besides the accuracy, the efficiency of an ANN also depends on the typical training time and how this time scales with the system size. The \textquotedblleft time\textquotedblright evolution of the accuracy during a training process is shown in Fig.\ref{fig:fig3} (b), where we observe that the ANN  first experiences a period of \textquotedblleft confusion\textquotedblright  with the prediction accuracy $P\simeq 0.5$ until a certain time $T_c$, after which the machine finally learns the correct approximation of the exact sign function and the accuracy will increase rapidly before saturating. This time scale $T_c$, together with the number of the neurons $N_b$, can be understood as the computation resources one needs to capture the sign rule using the ANN. The scaling relation between $T_c$ and the system size $L$ is plotted in the inset of Fig.\ref{fig:fig3} (b). The system size we studied is relatively small thus it is difficult to tell whether $T_c$ scales with L in a polynomial or exponential manner, which remains an open question.

Now we discuss more details of the ANN. First, even though throughout this paper we choose a ANN with a two-hidden layer structure, one may wonder whether further increasing the number of the layer can improve the prediction accuracy or not. To address this issue, we calculate $P$ for different ANN with different hidden layer (up to three) with fixed iteration steps.  As shown in Fig. \ref{fig:fig2} (c), we found that, at least for this example, even though an ANN with two hidden layers indeed performs much better than that with only one hidden layer, further increasing the number of layers does not significantly improve the performances. We also check the correlation between the erroneously assigned signs for a certain coefficient and the coefficient's absolute value. To do that, we define a parameter $\varepsilon$ to measure the ration between the average absolute value of the coefficient with the erroneously assigned sign and that of the whole training set.
\begin{equation}
\varepsilon=\frac{\frac 1{D'}\sum_{\boldsymbol{\sigma}'}|C[\boldsymbol{\sigma}']|}{\frac {1}{\mathbb{D}}\sum_{\boldsymbol{\sigma}}|C[\boldsymbol{\sigma}]|}
\end{equation}
where $[\boldsymbol{\sigma}']$ denotes the set of input basis with the erroneous assigned sign, with the dimensionality $D'$, while $[\boldsymbol{\sigma}]$ denotes all the training basis with the dimensionality ${\mathbb{D}}$.  As shown in Fig.\ref{fig:fig3} (d), we can find that $\varepsilon\sim\mathcal{O}(10^{-2})\ll 1$, indicating that the erroneous predictions of the sign-ANN tend to occur for those input basis whose coefficient with small absolute value.

 By combining the results of the amplitude (as shown in Fig.\ref{fig:fig3} (e) for examples) and sign ANNs, we can calculate the ground state of this frustrated quantum magnetic model, which agrees very well with the exact results, as shown in Fig.\ref{fig:fig3} (f). In a summary, by dividing the ANN into two part with different architecture, we can approximate the ground state of a frustrated quantum magnetism with high precisions. This strategy may shed light on using ANNs to solve the complex quantum many-body systems with sign problems.

\begin{figure}[htb]
\includegraphics[width=0.49\linewidth,bb=0 0 306 228]{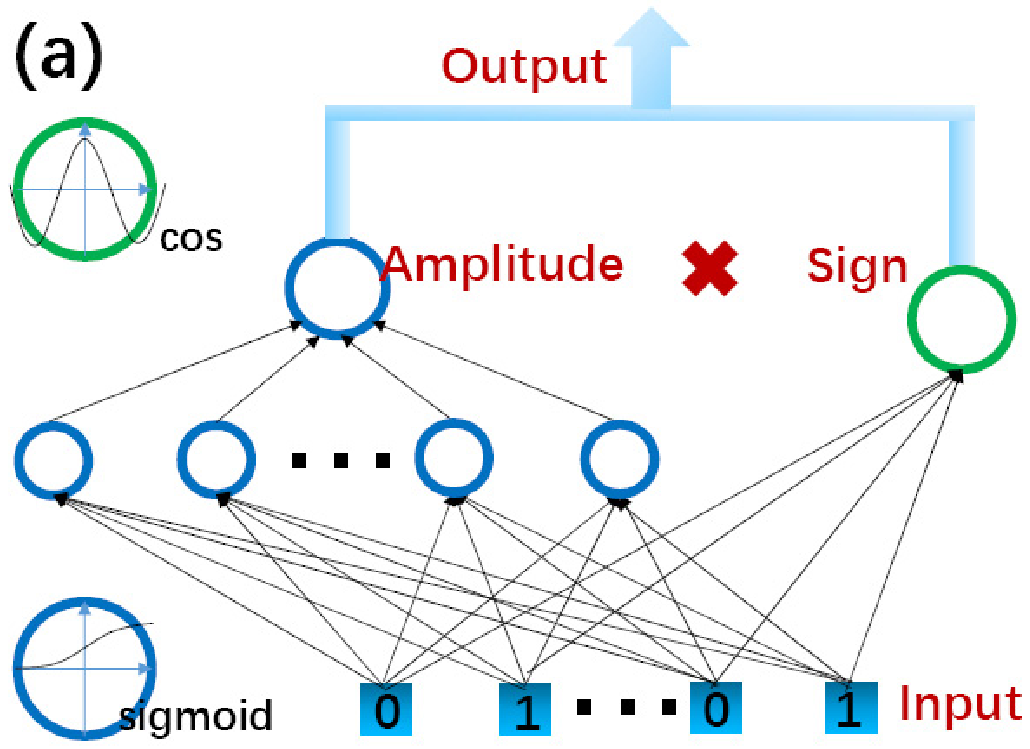}
\includegraphics[width=0.49\linewidth,bb=74 48 733 538]{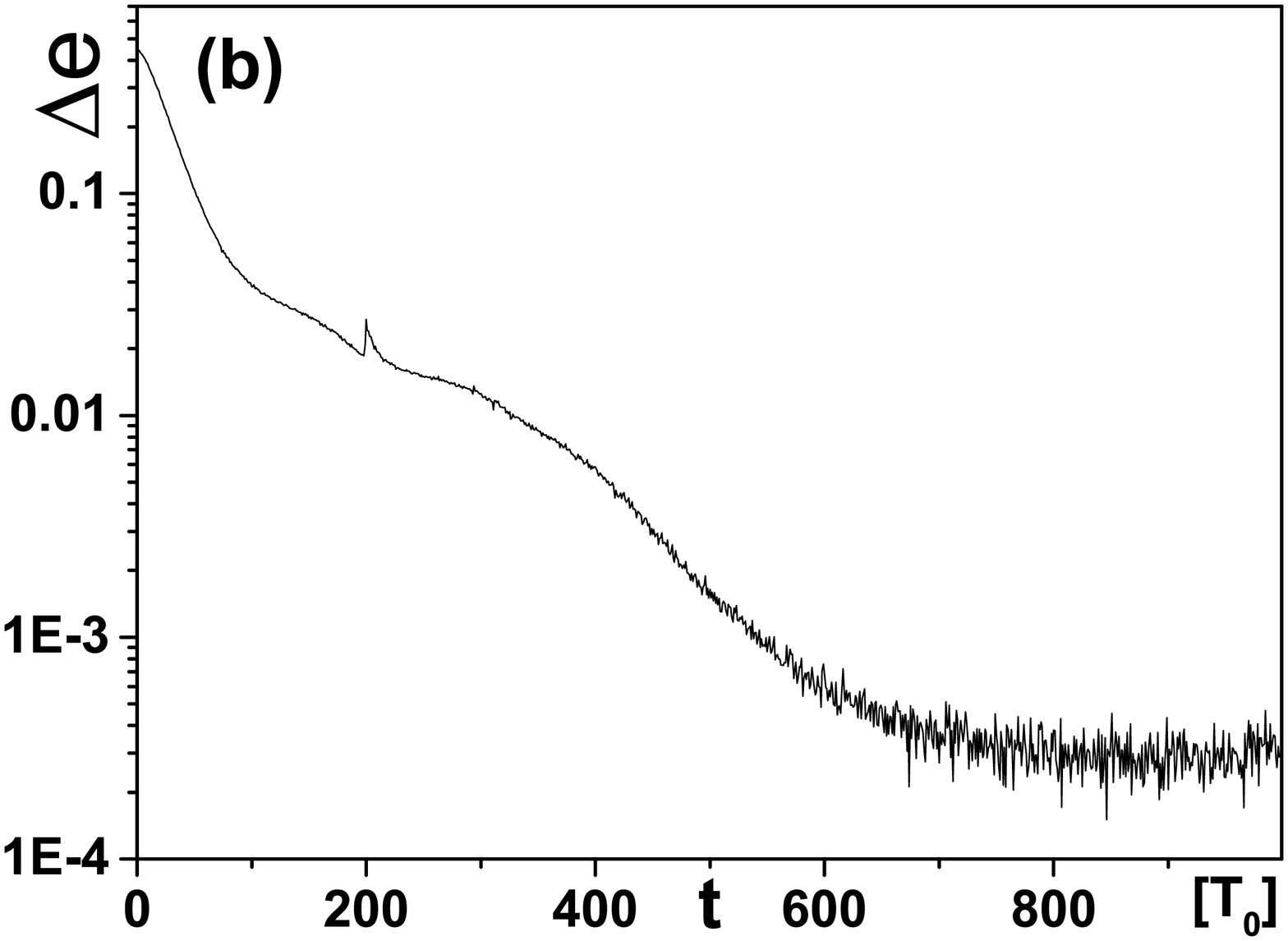}
\caption{ (a) The structure of the ANN we used in the variational method calculations. (b)the \textquotedblleft time\textquotedblright evolution of the excess energy per site $\Delta e=(E_v-E_0)/L$ during the training process for a 1D $J_1-J_2$ AF Heisenberg model with $L=30$  using the trial wave function as shown in Fig.\ref{fig:fig4} (a) with the neuron number in the hidden layer of the amplitude ANN $N_b=30$.}
\label{fig:fig4}
\end{figure}

\section {Variational results}
\label{sec:variational}

In all the cases studied above,  because the target functions (the ground state wave functions) are given, either  analytically or numerically, one may expect that the current method would not be useful for exploring new quantum many-body systems with previously unknown ground states. However, our previous results have established that the ground states of some quantum many-body systems can be efficiently represented by ANNs, which enables us to consider an ANN as a variational wave function for the true ground state of a new system. The connection weights and the bias in an ANN are the variational parameters with respect to which we seek to minimize the expectation value of the Hamiltonian(energy), instead of the fidelity function as described above.  For a given ANN with a set of parameters $\{W\}$, the corresponding variational energy  $E(\{W\})$ can be estimated using Monte Carlo simulations, with a  procedure similar to those used for previously studied free fermion cases in Sec.\ref{sec:free}. The minimum of $E(\{W\})$ in the space of parameters can be found using the stochastic reconfiguration optimization method. Similar strategy has been used in a different type to ANN: the restricted Boltzmann machine\cite{Carleo2017} to solve the non-frustrated quantum magnetic models e.g. the Heisenberg model and transverse Ising model. For the complex problems like the frustrated quantum magnetisms, we shall show that the strategy we proposed by dividing the ANNs into amplitude and sign parts can also improve the efficiency of this variational method.

In the following, we use the ANN  combined with variational Monte Carlo (VMC) methods (the details of the method can be found in the Appendix) to solve the ground state of an 1D $L=30$ $J_1-J_2$ AF Heisenberg model. The structure of the ANN we used in the variational method is shown in Fig. \ref{fig:fig4} (a). We adopt the strategy proposed in Sec.\ref{sec:Frustration} that the ANN has been divided into the amplitude and sign parts, and the final output is the product of them. However, one of the crucial differences is the output of the sign ANN is a continuous number instead of a discrete one. The reason is that a discrete function is usually non-differentiable, which may decrease the efficiency of the variational method.  We also notice that the structure of the ANN in \ref{fig:fig4} (a)  is much simpler than the one we used in Sec.\ref{sec:Frustration}:  there is only one hidden layer instead of two in both amplitude and sign ANN. We will turn back to this point later.

Since the target wave-function is unknown, the cost function to be minimized during the training process is the variational energy of the system, which can be evaluated by the VMC method for a given trial wave-function (ANN). In Fig.\ref{fig:fig4} (b), we plot the \textquotedblleft time\textquotedblright evolution of the excess energy per site $\Delta e=(E_v-E_0)/L$ during the training process, where $E_v$ is the variational energy evaluated by VMC and $E_0$ is the exact ground state energy calculated by Density Matrix Renormalization Group
method.  As shown in Fig. \ref{fig:fig4} (b), an ANN with a simple structure shown in Fig. \ref{fig:fig4} (a) can give rise to value of the ground state energy with precisions $\sim \mathcal{O}(10^{-3})$.

It is interesting to notice that compared to the ANN we used in Sec.\ref{sec:Frustration},   the structure of the ANN we used in this variational method is much simpler (only one hidden layer), but its performance is even better. There are several reasons for this counter-intuitive fact: (i) in the variational method, the sign and amplitude ANN are trained as a whole, while in Sec.\ref{sec:Frustration} the they are trained separately.  However, for the sign ANN, an extremely high accuracy may not be very necessary for the final results, since as shown in Fig.\ref{fig:fig3} (d) the erroneous predictions of the sign-ANN tend to occur for those input basis whose coefficient with very small absolute value, which give very little contribution to the final results, especially to the energy. (ii) In all the sections except Sec.\ref{sec:variational}, during the optimization process the training data sets are divided into $N$ batches and the optimization is performed  batch by batch, while in the variational method, all the entries in wave function are optimized simultaneously. (iii) In Sec.\ref{sec:Frustration}, we choose the cost function as the overlap of the wave functions (fidelity), which is a global quantity thus is usually difficult to be minimized, while in variational method, we only need to minimize a local quantity: the variational energy.

\section {Conclusion and Outlook}
In this paper, we demonstrated the powerful applicability of a simple ANN in approximating the ground state wavefunctions of some notable quantum many-body systems.  Even though an ANN with a simple structure can already approximate some of them with a high precision, there is still a long way to go before this method can learn to solve problems that remain inaccessible by any  other well-established numerical method, and the efficiency improvement plays a key role in this process.

Some avenues along this line  suggest the directions for future studies. First, as with many existing numerical methods,   imposing the Hamiltonian symmetries on the ANN will significantly improve its efficiency. Consider a 2D system for example.  An ANN with translational symmetry does not only reduce the number of the variational parameters and training data, but also learns the lattice geometry from the beginning. Inspired by the impressive success of deep learning techniques, we expect that an ANN can be more powerful when networks are made deeper.  In numerical simulations, we found that even though an ANN with two hidden layers indeed performs much better than that with only one hidden layer, further increasing the number of layers does not significantly improve the performances. One of the possible reason is that for an ANN with more hidden layers, even though its expressibility may be more powerful, the computational cost to find the optimal representation is significantly increased since the landscape in the parameter space is more complex.  Incorporation of \textquotedblleft deep learning\textquotedblright into our simulations remains an open issue and deserves further studies.

A fully connected ANN used in our simulations may contain many redundant connections that complicate the optimization process. Adopting  ANN with more advanced architectures({\it e.g.} convolutional neuron networks) may help to avoid this redundancy, thus likely significantly improve its efficiency. Recently, the relation of the quantum entanglement and the range of the connections in the ANN have been built, and we expect the convolutional neuron networks may work well for those ground state with short-range entanglement\cite{Deng2016b}.  Last but not least, it is known that the artificial neural network, in general,  is a heuristic algorithm whose efficiency largely depends on the designer's experience and intuition, however, to make this approach valuable in the physics, a systematic understanding of its validity and limitation,  at least for this concrete problem,  is still needed, which may be also beneficial to  the artificial intelligence community.

\appendix
\section{Details of the optimization algorithm}
During the training process of the artificial neural network, one needs to adjust the parameters in ANN to minimize the distance function, which turns to an multi-variable optimization problem. In our simulation, we use two optimization techniques depending on the concrete problems: the stochastic gradient descent (SGD)\cite{Goodfellow2016} and an adaptive learning rate optimization algorithm: adaptive moments (Adam). Here we only explain the details of the SGD method, and the Adam optimization algorithm has been explicitly illustrated in Ref.\cite{Kingma2014}.

In our simulation, each input data is composed of the basis and the corresponding coefficient (lable) calculated by ED or other methods (e.g. $\{ [\boldsymbol{\sigma}],C[\boldsymbol{\sigma}]\}$). For state with large Hilbert space dimension $\mathbb{D}$,  the typical coefficient is pretty small $C[\boldsymbol{\sigma}]\sim \mathcal{O}(1/\sqrt{\mathbb{D}})$. In general, the nonlinear activation function is not sensitive in the regime where the input/output is too small, therefore to make the ANN more efficient, we multiply the target function by a factor of $\sqrt{\mathbb{D}}$, thus $C[\boldsymbol{\sigma}]\sim \mathcal{O}(1)$. This renormalization doesn't change the results of any physical observable, but can significantly improve the efficiency of the fitting.   The input data set are randomly divided  into two groups: $80\%$ of them are used for training and the rest are testing data sets, and we choose a random set of ANN parameters $\mathbf{W}_0$ as the initial parameters. The training set of data are randomly reshuffled and divided into $N$ batches each of which containing $M$ data.  In each batch, the training data are labeled as $\{\boldsymbol{\sigma}^{(1)},\cdots\boldsymbol{\sigma}^{(M)}\}$ with corresponding target  $C[\boldsymbol{\sigma}^{(i)}]$.

After the initializations are finished, we start the optimization process. In each step of the SGD update, we  choose one batch and calculate the gradient estimate in the parameter landscape $\mathbf{W}$:
\begin{equation}
\mathbf{g}=\nabla_{\mathbf{W}}\frac 1M\sum_{i=1}^M L(C_p[\boldsymbol{\sigma}^{(i)}], C[\boldsymbol{\sigma}^{(i)}])
\end{equation}
where $C_p[\boldsymbol{\sigma}]$ is the coefficient predicted by the ANN and  $ L(C_p[\boldsymbol{\sigma}], C[\boldsymbol{\sigma}])=(C[\boldsymbol{\sigma}]- C_p[\boldsymbol{\sigma}])^2$ is a function of $\mathbf{W}$, denoted as loss function.  Once we obtain the gradient, the parameter $\mathbf{W}$ is updated as:
\begin{equation}
\mathbf{W}\leftarrow \mathbf{W}-\epsilon\mathbf{g}
\end{equation}
where $\epsilon$ is the parameter controlling the learning rate, which gradually decreases over time. The above optimization processes continue until all the batches are chosen, then one iteration of the training is finished. In our simulations, the training time is measured in the unit of the time of a single iteration $T_0$. The typical training time range from $10^2\sim 10^3$ $T_0$ depending on the convergency of the problems.   The SGD method is most used optimization technique for machine learning. In our simulation, we use the GPU to speed up the computational efficiency of the training.

\section{Details of the variational method}

In this section, we will show the variational analysis of using the artificial neural network to explore new ground states. In this case, the target function is previously unknown, therefore, during the training process,  what we need to minimize is not the distance function $F$, but the variational energy (the expectation value of the Hamiltonian over the ANN trial wave function), which can be calculated using the variational Monte Carlo method, as we will show in the following. In general, a Hamiltonian can be split into the diagonal and non-diagonal parts: $\hat{H}=\hat{T}+\hat{V}$ in the  basis $|\boldsymbol{\sigma}\rangle$ (or Fock basis). For a given trial wave function in terms of the ANN with a set of  variational parameter $\{W\}$,  the expectation value of the diagonal part:
\begin{equation}
V(\{W\})=\langle \hat{V}\rangle=\sum_{[\boldsymbol{\sigma}]}p_{\boldsymbol{\sigma}}\{W\} \langle \boldsymbol{\sigma}|\hat{V}|\boldsymbol{\sigma}\rangle
\end{equation}
where $p_{\boldsymbol{\sigma}}\{W\}=|C_{ANN}[\boldsymbol{\sigma}]|^2$ is the probability predicted by the ANN. The summation  $\sum_{[\boldsymbol{\sigma}]}$ is over the whole Hilbert space, whose dimensionality exponentially grows with the system size. Follow the spirt of Monte Carlo, the summation over the whole Hilbert space can be replaced by the summation over those important configurations:
\begin{equation}
V(\{W\})\simeq \sum_{[\boldsymbol{\bar{\sigma}}]}\frac{p_{\boldsymbol{\bar{\sigma}}}\{W\}}{Z\{W\}} \langle \boldsymbol{\bar{\sigma}}|\hat{V}|\boldsymbol{\bar{\sigma}}\rangle \label{eq:V}
\end{equation}
with $Z\{W\}=\sum_{[\boldsymbol{\bar{\sigma}}]}p_{\boldsymbol{\bar{\sigma}}}\{W\}$, and $[\bar{\boldsymbol{\sigma}}]$ is an exponentially small fraction of the whole configurations, denotes the set of representative configurations chosen by the importance sampling using Metropolis algorithm according to its probability $p_{\boldsymbol{\bar{\sigma}}}$. The expectation value of the off-diagonal term can be evaluated in a similar way, without loss of generality, we assume $T=|\boldsymbol{\sigma'}\rangle\langle\boldsymbol{\sigma}|$ with $|\boldsymbol{\sigma}\rangle\neq|\boldsymbol{\sigma'}\rangle$, thus
\begin{equation}
T(\{W\})\simeq \sum_{[\boldsymbol{\bar{\sigma}}]}\frac{p_{\boldsymbol{\bar{\sigma}}}\{W\}}{Z\{W\}}\times \frac{C_{ANN}[\boldsymbol{\bar{\sigma'}}]}{C_{ANN}[\boldsymbol{\bar{\sigma}}]}\langle \boldsymbol{\bar{\sigma'}}|\hat{T}|\boldsymbol{\bar{\sigma}}\rangle \label{eq:T}
\end{equation}
where the observable we need to calculate during the sampling is not only $\langle\boldsymbol{\bar{\sigma'}}|\hat{T}|\boldsymbol{\bar{\sigma}}\rangle$, but $\langle \boldsymbol{\bar{\sigma'}}|\hat{T}|\boldsymbol{\bar{\sigma}}\rangle C_{ANN}[\boldsymbol{\bar{\sigma'}}]/C_{ANN}[\boldsymbol{\bar{\sigma}}]$. By combining Eq.(\ref{eq:V}) and Eq.(\ref{eq:T}), we can obtain the variational energy for a given trail ANN wave function. To minimize the variational energy, one needs to calculate the derivative of $H(\{W\})$ with respect of $\{W\}$, and the variational parameters can be determined by solving the equation:
\begin{equation}
\frac{\partial H(\{W\})}{\partial w_i}=\frac{\partial V(\{W\})}{\partial w_i}+\frac{\partial T(\{W\})}{\partial w_i}=0 \label{eq:equation}
\end{equation}
Notice that an ANN is a combination of a set of non-linear functions whose explicitly forms are already known, as a consequence, one can easily obtain its derivative $\frac {\partial C^{\{W\}}_{ANN}[\boldsymbol{\sigma}]}{\partial w_i}$ , thus:
\begin{eqnarray}
\nonumber \frac{\partial V(\{W\})}{\partial w_i}&=&\sum_{[\boldsymbol{\sigma}]}p_{\boldsymbol{\sigma}}\{W\} \times \frac {\partial C^{\{W\}}_{ANN}[\boldsymbol{\sigma}]}{\partial w_i} \frac{2\langle \boldsymbol{\sigma}|\hat{V}|\boldsymbol{\sigma}\rangle}{C^{\{W\}}_{ANN}[\boldsymbol{\sigma}]}\\
\nonumber \frac{\partial T(\{W\})}{\partial w_i}&=&\sum_{[\boldsymbol{\sigma}]}p_{\boldsymbol{\sigma}}\{W\} \times \frac{\langle \boldsymbol{\sigma'}|\hat{T}|\boldsymbol{\sigma}\rangle}{C^{\{W\}}_{ANN}[\boldsymbol{\sigma}]}  \\
\nonumber &\times& [\frac {\partial C^{\{W\}}_{ANN}[\boldsymbol{\sigma'}]}{\partial w_i}+ \frac {\partial C^{\{W\}}_{ANN}[\boldsymbol{\sigma}]}{\partial w_i}\frac{C^{\{W\}}_{ANN}[\boldsymbol{\sigma'}]}{C^{\{W\}}_{ANN}[\boldsymbol{\sigma}]}]
\end{eqnarray}
By performing the importance sampling, one can calculate the $\frac{\partial V(\{W\})}{\partial w_i}$ and $\frac{\partial T(\{W\})}{\partial w_i}$, then substitute them into Eq.(\ref{eq:equation}) to solve the variational parameters.

{\it Acknowledgement--}We wish to thank Lei Wang and J. Carrasquilla for fruitful discussions. ZC is supported by the National Key Research and Development Program of China (grant No. 2016YFA0302001), the National Natural Science Foundation of China under Grant No.11674221 and the Shanghai Rising-Star Program. J.G.L. is supported by the National Natural Science Foundation of China under Grant No. 11774398.


\begin{thebibliography}{42}%
\makeatletter
\providecommand \@ifxundefined [1]{%
 \@ifx{#1\undefined}
}%
\providecommand \@ifnum [1]{%
 \ifnum #1\expandafter \@firstoftwo
 \else \expandafter \@secondoftwo
 \fi
}%
\providecommand \@ifx [1]{%
 \ifx #1\expandafter \@firstoftwo
 \else \expandafter \@secondoftwo
 \fi
}%
\providecommand \natexlab [1]{#1}%
\providecommand \enquote  [1]{``#1''}%
\providecommand \bibnamefont  [1]{#1}%
\providecommand \bibfnamefont [1]{#1}%
\providecommand \citenamefont [1]{#1}%
\providecommand \href@noop [0]{\@secondoftwo}%
\providecommand \href [0]{\begingroup \@sanitize@url \@href}%
\providecommand \@href[1]{\@@startlink{#1}\@@href}%
\providecommand \@@href[1]{\endgroup#1\@@endlink}%
\providecommand \@sanitize@url [0]{\catcode `\\12\catcode `\$12\catcode
  `\&12\catcode `\#12\catcode `\^12\catcode `\_12\catcode `\%12\relax}%
\providecommand \@@startlink[1]{}%
\providecommand \@@endlink[0]{}%
\providecommand \url  [0]{\begingroup\@sanitize@url \@url }%
\providecommand \@url [1]{\endgroup\@href {#1}{\urlprefix }}%
\providecommand \urlprefix  [0]{URL }%
\providecommand \Eprint [0]{\href }%
\providecommand \doibase [0]{http://dx.doi.org/}%
\providecommand \selectlanguage [0]{\@gobble}%
\providecommand \bibinfo  [0]{\@secondoftwo}%
\providecommand \bibfield  [0]{\@secondoftwo}%
\providecommand \translation [1]{[#1]}%
\providecommand \BibitemOpen [0]{}%
\providecommand \bibitemStop [0]{}%
\providecommand \bibitemNoStop [0]{.\EOS\space}%
\providecommand \EOS [0]{\spacefactor3000\relax}%
\providecommand \BibitemShut  [1]{\csname bibitem#1\endcsname}%
\let\auto@bib@innerbib\@empty
\bibitem [{\citenamefont {White}(1992)}]{White1992}%
  \BibitemOpen
  \bibfield  {author} {\bibinfo {author} {\bibfnamefont {Steven~R.}\
  \bibnamefont {White}},\ }\bibfield  {title} {\enquote {\bibinfo {title}
  {Density matrix formulation for quantum renormalization groups},}\
  }\href@noop {} {\bibfield  {journal} {\bibinfo  {journal} {Phys. Rev. Lett.}\
  }\textbf {\bibinfo {volume} {69}},\ \bibinfo {pages} {2863--2866} (\bibinfo
  {year} {1992})}\BibitemShut {NoStop}%
\bibitem [{\citenamefont {Verstraete}\ \emph {et~al.}(2008)\citenamefont
  {Verstraete}, \citenamefont {Murg},\ and\ \citenamefont
  {Cirac}}]{Verstraete2008}%
  \BibitemOpen
  \bibfield  {author} {\bibinfo {author} {\bibfnamefont {F.}~\bibnamefont
  {Verstraete}}, \bibinfo {author} {\bibfnamefont {V.}~\bibnamefont {Murg}}, \
  and\ \bibinfo {author} {\bibfnamefont {J.I.}\ \bibnamefont {Cirac}},\
  }\bibfield  {title} {\enquote {\bibinfo {title} {Matrix product states,
  projected entangled pair states, and variational renormalization group
  methods for quantum spin systems},}\ }\href@noop {} {\bibfield  {journal}
  {\bibinfo  {journal} {Advances in Physics}\ }\textbf {\bibinfo {volume}
  {57}},\ \bibinfo {pages} {143--224} (\bibinfo {year} {2008})}\BibitemShut
  {NoStop}%
\bibitem [{\citenamefont {Schollwoeck}(2011)}]{Schollwock2011}%
  \BibitemOpen
  \bibfield  {author} {\bibinfo {author} {\bibfnamefont {Ulrich}\ \bibnamefont
  {Schollwoeck}},\ }\bibfield  {title} {\enquote {\bibinfo {title} {The
  density-matrix renormalization group in the age of matrix product states},}\
  }\href@noop {} {\bibfield  {journal} {\bibinfo  {journal} {Annals of
  Physics}\ }\textbf {\bibinfo {volume} {326}},\ \bibinfo {pages} {96 -- 192}
  (\bibinfo {year} {2011})}\BibitemShut {NoStop}%
\bibitem [{\citenamefont {Sandvik}\ and\ \citenamefont
  {Kurkij\"arvi}(1991)}]{Sandvik1991}%
  \BibitemOpen
  \bibfield  {author} {\bibinfo {author} {\bibfnamefont {Anders~W.}\
  \bibnamefont {Sandvik}}\ and\ \bibinfo {author} {\bibfnamefont {Juhani}\
  \bibnamefont {Kurkij\"arvi}},\ }\bibfield  {title} {\enquote {\bibinfo
  {title} {Quantum monte carlo simulation method for spin systems},}\
  }\href@noop {} {\bibfield  {journal} {\bibinfo  {journal} {Phys. Rev. B}\
  }\textbf {\bibinfo {volume} {43}},\ \bibinfo {pages} {5950--5961} (\bibinfo
  {year} {1991})}\BibitemShut {NoStop}%
\bibitem [{\citenamefont {Prokof'ev}\ \emph {et~al.}(1998)\citenamefont
  {Prokof'ev}, \citenamefont {Svistunov},\ and\ \citenamefont
  {Tupitsyn}}]{Prokofev1998}%
  \BibitemOpen
  \bibfield  {author} {\bibinfo {author} {\bibfnamefont {N.~V.}\ \bibnamefont
  {Prokof'ev}}, \bibinfo {author} {\bibfnamefont {B.~V.}\ \bibnamefont
  {Svistunov}}, \ and\ \bibinfo {author} {\bibfnamefont {I.~S.}\ \bibnamefont
  {Tupitsyn}},\ }\href@noop {} {\bibfield  {journal} {\bibinfo  {journal}
  {Phys. Lett. A}\ }\textbf {\bibinfo {volume} {238}},\ \bibinfo {pages} {253}
  (\bibinfo {year} {1998})}\BibitemShut {NoStop}%
\bibitem [{\citenamefont {Gubernatis}\ \emph {et~al.}(2016)\citenamefont
  {Gubernatis}, \citenamefont {Kawashima},\ and\ \citenamefont
  {Werner}}]{Gubernatis2016}%
  \BibitemOpen
  \bibfield  {author} {\bibinfo {author} {\bibfnamefont {J.}~\bibnamefont
  {Gubernatis}}, \bibinfo {author} {\bibfnamefont {N.}~\bibnamefont
  {Kawashima}}, \ and\ \bibinfo {author} {\bibfnamefont {P.}~\bibnamefont
  {Werner}},\ }\href@noop {} {\emph {\bibinfo {title} {Quantum Monte Carlo
  Methods: Algorithms for Lattice Models}}}\ (\bibinfo  {publisher} {~Cambridge
  University Press, Cambridge},\ \bibinfo {year} {2016})\BibitemShut {NoStop}%
\bibitem [{\citenamefont {Nielsen}(2015)}]{Nielsen2015}%
  \BibitemOpen
  \bibfield  {author} {\bibinfo {author} {\bibfnamefont {M.~A.}\ \bibnamefont
  {Nielsen}},\ }\href@noop {} {\emph {\bibinfo {title} {Neural Networks and
  Deep Learning}}}\ (\bibinfo  {publisher} {~Determination Press},\ \bibinfo
  {year} {2015})\BibitemShut {NoStop}%
\bibitem [{\citenamefont {Goodfellow}\ \emph {et~al.}(2016)\citenamefont
  {Goodfellow}, \citenamefont {Bengio},\ and\ \citenamefont
  {Courville}}]{Goodfellow2016}%
  \BibitemOpen
  \bibfield  {author} {\bibinfo {author} {\bibfnamefont {I.}~\bibnamefont
  {Goodfellow}}, \bibinfo {author} {\bibfnamefont {Y.}~\bibnamefont {Bengio}},
  \ and\ \bibinfo {author} {\bibfnamefont {A.}~\bibnamefont {Courville}},\
  }\href@noop {} {\emph {\bibinfo {title} {Deep learning}}}\ (\bibinfo
  {publisher} {~book in preparation for MIT Press},\ \bibinfo {year}
  {2016})\BibitemShut {NoStop}%
\bibitem [{\citenamefont {{Mehta}}\ and\ \citenamefont
  {{Schwab}}(2014)}]{Mehta2014}%
  \BibitemOpen
  \bibfield  {author} {\bibinfo {author} {\bibfnamefont {P.}~\bibnamefont
  {{Mehta}}}\ and\ \bibinfo {author} {\bibfnamefont {D.~J.}\ \bibnamefont
  {{Schwab}}},\ }\bibfield  {title} {\enquote {\bibinfo {title} {{An exact
  mapping between the Variational Renormalization Group and Deep Learning}},}\
  }\href@noop {} {\bibfield  {journal} {\bibinfo  {journal} {ArXiv e-prints}\ }
  (\bibinfo {year} {2014})},\ \Eprint {http://arxiv.org/abs/1410.3831}
  {arXiv:1410.3831 [stat.ML]} \BibitemShut {NoStop}%
\bibitem [{\citenamefont {{Lin}}\ and\ \citenamefont
  {{Tegmark}}(2017)}]{Lin2016}%
  \BibitemOpen
  \bibfield  {author} {\bibinfo {author} {\bibfnamefont {H.~W.}\ \bibnamefont
  {{Lin}}}\ and\ \bibinfo {author} {\bibfnamefont {M.}~\bibnamefont
  {{Tegmark}}},\ }\bibfield  {title} {\enquote {\bibinfo {title} {{Why does
  deep and cheap learning work so well?}}}\ }\href@noop {} {\bibfield
  {journal} {\bibinfo  {journal} {J. Stat. Phys}\ }\textbf {\bibinfo {volume}
  {168}},\ \bibinfo {pages} {1223} (\bibinfo {year} {2017})}\BibitemShut
  {NoStop}%
\bibitem [{\citenamefont {{Chen}}\ \emph {et~al.}(2017)\citenamefont {{Chen}},
  \citenamefont {{Cheng}}, \citenamefont {{Xie}}, \citenamefont {{Wang}},\ and\
  \citenamefont {{Xiang}}}]{Chen2017}%
  \BibitemOpen
  \bibfield  {author} {\bibinfo {author} {\bibfnamefont {J.}~\bibnamefont
  {{Chen}}}, \bibinfo {author} {\bibfnamefont {S.}~\bibnamefont {{Cheng}}},
  \bibinfo {author} {\bibfnamefont {H.}~\bibnamefont {{Xie}}}, \bibinfo
  {author} {\bibfnamefont {L.}~\bibnamefont {{Wang}}}, \ and\ \bibinfo {author}
  {\bibfnamefont {T.}~\bibnamefont {{Xiang}}},\ }\bibfield  {title} {\enquote
  {\bibinfo {title} {{On the Equivalence of Restricted Boltzmann Machines and
  Tensor Network States}},}\ }\href@noop {} {\bibfield  {journal} {\bibinfo
  {journal} {ArXiv e-prints}\ } (\bibinfo {year} {2017})},\ \Eprint
  {http://arxiv.org/abs/1701.04831} {arXiv:1701.04831 [cond-mat.str-el]}
  \BibitemShut {NoStop}%
\bibitem [{\citenamefont {Stoudenmire}\ and\ \citenamefont
  {Schwab}(2016)}]{Stoudenmire2016}%
  \BibitemOpen
  \bibfield  {author} {\bibinfo {author} {\bibfnamefont {E.~Miles}\
  \bibnamefont {Stoudenmire}}\ and\ \bibinfo {author} {\bibfnamefont
  {David~J.}\ \bibnamefont {Schwab}},\ }\href@noop {} {\bibfield  {journal}
  {\bibinfo  {journal} {Advances in Neural Information Processing Systems}\
  }\textbf {\bibinfo {volume} {29}},\ \bibinfo {pages} {4799} (\bibinfo {year}
  {2016})}\BibitemShut {NoStop}%
\bibitem [{\citenamefont {Carrasquilla}\ and\ \citenamefont
  {Melko}(2017)}]{Carrasquilla2017}%
  \BibitemOpen
  \bibfield  {author} {\bibinfo {author} {\bibfnamefont {Juan}\ \bibnamefont
  {Carrasquilla}}\ and\ \bibinfo {author} {\bibfnamefont {Roger~G.}\
  \bibnamefont {Melko}},\ }\bibfield  {title} {\enquote {\bibinfo {title}
  {Machine learning phases of matter},}\ }\href@noop {} {\bibfield  {journal}
  {\bibinfo  {journal} {Nat Phys}\ }\textbf {\bibinfo {volume} {13}},\ \bibinfo
  {pages} {431} (\bibinfo {year} {2017})}\BibitemShut {NoStop}%
\bibitem [{\citenamefont {Wang}(2016)}]{Wang2016}%
  \BibitemOpen
  \bibfield  {author} {\bibinfo {author} {\bibfnamefont {Lei}\ \bibnamefont
  {Wang}},\ }\bibfield  {title} {\enquote {\bibinfo {title} {Discovering phase
  transitions with unsupervised learning},}\ }\href@noop {} {\bibfield
  {journal} {\bibinfo  {journal} {Phys. Rev. B}\ }\textbf {\bibinfo {volume}
  {94}},\ \bibinfo {pages} {195105} (\bibinfo {year} {2016})}\BibitemShut
  {NoStop}%
\bibitem [{\citenamefont {van Nieuwenburg}\ \emph {et~al.}(2017)\citenamefont
  {van Nieuwenburg}, \citenamefont {Liu},\ and\ \citenamefont
  {Huber}}]{Nieuwenburg2017}%
  \BibitemOpen
  \bibfield  {author} {\bibinfo {author} {\bibfnamefont {Evert P.~L.}\
  \bibnamefont {van Nieuwenburg}}, \bibinfo {author} {\bibfnamefont {Ye-Hua}\
  \bibnamefont {Liu}}, \ and\ \bibinfo {author} {\bibfnamefont {Sebastian~D.}\
  \bibnamefont {Huber}},\ }\bibfield  {title} {\enquote {\bibinfo {title}
  {Learning phase transitions by confusion},}\ }\href@noop {} {\bibfield
  {journal} {\bibinfo  {journal} {Nat Phys}\ }\textbf {\bibinfo {volume}
  {13}},\ \bibinfo {pages} {435} (\bibinfo {year} {2017})}\BibitemShut
  {NoStop}%
\bibitem [{\citenamefont {Deng}\ \emph {et~al.}(2017)\citenamefont {Deng},
  \citenamefont {Li},\ and\ \citenamefont {Das~Sarma}}]{Deng2016b}%
  \BibitemOpen
  \bibfield  {author} {\bibinfo {author} {\bibfnamefont {Dong-Ling}\
  \bibnamefont {Deng}}, \bibinfo {author} {\bibfnamefont {Xiaopeng}\
  \bibnamefont {Li}}, \ and\ \bibinfo {author} {\bibfnamefont {S.}~\bibnamefont
  {Das~Sarma}},\ }\bibfield  {title} {\enquote {\bibinfo {title} {Quantum
  entanglement in neural network states},}\ }\href@noop {} {\bibfield
  {journal} {\bibinfo  {journal} {Phys. Rev. X}\ }\textbf {\bibinfo {volume}
  {7}},\ \bibinfo {pages} {021021} (\bibinfo {year} {2017})}\BibitemShut
  {NoStop}%
\bibitem [{\citenamefont {{Torlai}}\ \emph {et~al.}(2017)\citenamefont
  {{Torlai}}, \citenamefont {{Mazzola}}, \citenamefont {{Carrasquilla}},
  \citenamefont {{Troyer}}, \citenamefont {{Melko}},\ and\ \citenamefont
  {{Carleo}}}]{Torlai2017}%
  \BibitemOpen
  \bibfield  {author} {\bibinfo {author} {\bibfnamefont {G.}~\bibnamefont
  {{Torlai}}}, \bibinfo {author} {\bibfnamefont {G.}~\bibnamefont {{Mazzola}}},
  \bibinfo {author} {\bibfnamefont {J.}~\bibnamefont {{Carrasquilla}}},
  \bibinfo {author} {\bibfnamefont {M.}~\bibnamefont {{Troyer}}}, \bibinfo
  {author} {\bibfnamefont {R.}~\bibnamefont {{Melko}}}, \ and\ \bibinfo
  {author} {\bibfnamefont {G.}~\bibnamefont {{Carleo}}},\ }\bibfield  {title}
  {\enquote {\bibinfo {title} {{Many-body quantum state tomography with neural
  networks}},}\ }\href@noop {} {\bibfield  {journal} {\bibinfo  {journal}
  {ArXiv e-prints}\ } (\bibinfo {year} {2017})},\ \Eprint
  {http://arxiv.org/abs/1703.05334} {arXiv:1703.05334 [cond-mat.dis-nn]}
  \BibitemShut {NoStop}%
\bibitem [{\citenamefont {{Ohtsuki}}\ and\ \citenamefont
  {{Ohtsuki}}(2016)}]{Ohtsuki2016}%
  \BibitemOpen
  \bibfield  {author} {\bibinfo {author} {\bibfnamefont {T.}~\bibnamefont
  {{Ohtsuki}}}\ and\ \bibinfo {author} {\bibfnamefont {T.}~\bibnamefont
  {{Ohtsuki}}},\ }\bibfield  {title} {\enquote {\bibinfo {title} {{Deep
  Learning the Quantum Phase Transitions in Random Two-Dimensional Electron
  Systems }},}\ }\href@noop {} {\bibfield  {journal} {\bibinfo  {journal} {J.
  Phys. Soc. Jpn}\ }\textbf {\bibinfo {volume} {85}},\ \bibinfo {pages}
  {123706} (\bibinfo {year} {2016})}\BibitemShut {NoStop}%
\bibitem [{\citenamefont {Zhang}\ and\ \citenamefont {Kim}(2017)}]{Zhang2016}%
  \BibitemOpen
  \bibfield  {author} {\bibinfo {author} {\bibfnamefont {Yi}~\bibnamefont
  {Zhang}}\ and\ \bibinfo {author} {\bibfnamefont {Eun-Ah}\ \bibnamefont
  {Kim}},\ }\bibfield  {title} {\enquote {\bibinfo {title} {Quantum loop
  topography for machine learning},}\ }\href@noop {} {\bibfield  {journal}
  {\bibinfo  {journal} {Phys. Rev. Lett.}\ }\textbf {\bibinfo {volume} {118}},\
  \bibinfo {pages} {216401} (\bibinfo {year} {2017})}\BibitemShut {NoStop}%
\bibitem [{\citenamefont {Ch'ng}\ \emph {et~al.}(2017)\citenamefont {Ch'ng},
  \citenamefont {Carrasquilla}, \citenamefont {Melko},\ and\ \citenamefont
  {Khatami}}]{Ch2016}%
  \BibitemOpen
  \bibfield  {author} {\bibinfo {author} {\bibfnamefont {Kelvin}\ \bibnamefont
  {Ch'ng}}, \bibinfo {author} {\bibfnamefont {Juan}\ \bibnamefont
  {Carrasquilla}}, \bibinfo {author} {\bibfnamefont {Roger~G.}\ \bibnamefont
  {Melko}}, \ and\ \bibinfo {author} {\bibfnamefont {Ehsan}\ \bibnamefont
  {Khatami}},\ }\bibfield  {title} {\enquote {\bibinfo {title} {Machine
  learning phases of strongly correlated fermions},}\ }\href@noop {} {\bibfield
   {journal} {\bibinfo  {journal} {Phys. Rev. X}\ }\textbf {\bibinfo {volume}
  {7}},\ \bibinfo {pages} {031038} (\bibinfo {year} {2017})}\BibitemShut
  {NoStop}%
\bibitem [{\citenamefont {{Ponte}}\ and\ \citenamefont
  {{Melko}}(2017)}]{Ponte2017}%
  \BibitemOpen
  \bibfield  {author} {\bibinfo {author} {\bibfnamefont {P.}~\bibnamefont
  {{Ponte}}}\ and\ \bibinfo {author} {\bibfnamefont {R.~G.}\ \bibnamefont
  {{Melko}}},\ }\bibfield  {title} {\enquote {\bibinfo {title} {{Kernel methods
  for interpretable machine learning of order parameters}},}\ }\href@noop {}
  {\bibfield  {journal} {\bibinfo  {journal} {ArXiv e-prints}\ } (\bibinfo
  {year} {2017})},\ \Eprint {http://arxiv.org/abs/1704.05848} {arXiv:1704.05848
  [cond-mat.stat-mech]} \BibitemShut {NoStop}%
\bibitem [{\citenamefont {Curtarolo}\ \emph {et~al.}(2003)\citenamefont
  {Curtarolo}, \citenamefont {Morgan}, \citenamefont {Persson}, \citenamefont
  {Rodgers},\ and\ \citenamefont {Ceder}}]{Curtarolo2003}%
  \BibitemOpen
  \bibfield  {author} {\bibinfo {author} {\bibfnamefont {Stefano}\ \bibnamefont
  {Curtarolo}}, \bibinfo {author} {\bibfnamefont {Dane}\ \bibnamefont
  {Morgan}}, \bibinfo {author} {\bibfnamefont {Kristin}\ \bibnamefont
  {Persson}}, \bibinfo {author} {\bibfnamefont {John}\ \bibnamefont {Rodgers}},
  \ and\ \bibinfo {author} {\bibfnamefont {Gerbrand}\ \bibnamefont {Ceder}},\
  }\bibfield  {title} {\enquote {\bibinfo {title} {Predicting crystal
  structures with data mining of quantum calculations},}\ }\href@noop {}
  {\bibfield  {journal} {\bibinfo  {journal} {Phys. Rev. Lett.}\ }\textbf
  {\bibinfo {volume} {91}},\ \bibinfo {pages} {135503} (\bibinfo {year}
  {2003})}\BibitemShut {NoStop}%
\bibitem [{\citenamefont {Kalinin}\ \emph {et~al.}(2015)\citenamefont
  {Kalinin}, \citenamefont {Sumpter},\ and\ \citenamefont
  {Archibald}}]{Kalinin2015}%
  \BibitemOpen
  \bibfield  {author} {\bibinfo {author} {\bibfnamefont {S.~V.}\ \bibnamefont
  {Kalinin}}, \bibinfo {author} {\bibfnamefont {B.~G.}\ \bibnamefont
  {Sumpter}}, \ and\ \bibinfo {author} {\bibfnamefont {R.~K.}\ \bibnamefont
  {Archibald}},\ }\bibfield  {title} {\enquote {\bibinfo {title}
  {Big¨cdeep¨csmart data in imaging for guiding materials design},}\
  }\href@noop {} {\bibfield  {journal} {\bibinfo  {journal} {Nat Mater}\
  }\textbf {\bibinfo {volume} {14}},\ \bibinfo {pages} {973} (\bibinfo {year}
  {2015})}\BibitemShut {NoStop}%
\bibitem [{\citenamefont {Ghiringhelli}\ \emph {et~al.}(2015)\citenamefont
  {Ghiringhelli}, \citenamefont {Vybiral}, \citenamefont {Levchenko},
  \citenamefont {Draxl},\ and\ \citenamefont {Scheffler}}]{Ghiringhelli2015}%
  \BibitemOpen
  \bibfield  {author} {\bibinfo {author} {\bibfnamefont {Luca~M.}\ \bibnamefont
  {Ghiringhelli}}, \bibinfo {author} {\bibfnamefont {Jan}\ \bibnamefont
  {Vybiral}}, \bibinfo {author} {\bibfnamefont {Sergey~V.}\ \bibnamefont
  {Levchenko}}, \bibinfo {author} {\bibfnamefont {Claudia}\ \bibnamefont
  {Draxl}}, \ and\ \bibinfo {author} {\bibfnamefont {Matthias}\ \bibnamefont
  {Scheffler}},\ }\bibfield  {title} {\enquote {\bibinfo {title} {Big data of
  materials science: Critical role of the descriptor},}\ }\href@noop {}
  {\bibfield  {journal} {\bibinfo  {journal} {Phys. Rev. Lett.}\ }\textbf
  {\bibinfo {volume} {114}},\ \bibinfo {pages} {105503} (\bibinfo {year}
  {2015})}\BibitemShut {NoStop}%
\bibitem [{\citenamefont {{Broecker}}\ \emph {et~al.}(2017)\citenamefont
  {{Broecker}}, \citenamefont {{Carrasquilla}}, \citenamefont {{Melko}},\ and\
  \citenamefont {{Trebst}}}]{Broecker2016}%
  \BibitemOpen
  \bibfield  {author} {\bibinfo {author} {\bibfnamefont {P.}~\bibnamefont
  {{Broecker}}}, \bibinfo {author} {\bibfnamefont {J.}~\bibnamefont
  {{Carrasquilla}}}, \bibinfo {author} {\bibfnamefont {R.~G.}\ \bibnamefont
  {{Melko}}}, \ and\ \bibinfo {author} {\bibfnamefont {S.}~\bibnamefont
  {{Trebst}}},\ }\bibfield  {title} {\enquote {\bibinfo {title} {{Machine
  learning quantum phases of matter beyond the fermion sign problem}},}\
  }\href@noop {} {\bibfield  {journal} {\bibinfo  {journal} {Scientific
  Reports}\ }\textbf {\bibinfo {volume} {7}},\ \bibinfo {pages} {8823}
  (\bibinfo {year} {2017})}\BibitemShut {NoStop}%
\bibitem [{\citenamefont {Huang}\ \emph {et~al.}(2017)\citenamefont {Huang},
  \citenamefont {Yang},\ and\ \citenamefont {Wang}}]{Huang2016}%
  \BibitemOpen
  \bibfield  {author} {\bibinfo {author} {\bibfnamefont {Li}~\bibnamefont
  {Huang}}, \bibinfo {author} {\bibfnamefont {Yi-feng}\ \bibnamefont {Yang}}, \
  and\ \bibinfo {author} {\bibfnamefont {Lei}\ \bibnamefont {Wang}},\
  }\bibfield  {title} {\enquote {\bibinfo {title} {Recommender engine for
  continuous-time quantum monte carlo methods},}\ }\href@noop {} {\bibfield
  {journal} {\bibinfo  {journal} {Phys. Rev. E}\ }\textbf {\bibinfo {volume}
  {95}},\ \bibinfo {pages} {031301} (\bibinfo {year} {2017})}\BibitemShut
  {NoStop}%
\bibitem [{\citenamefont {Huang}\ and\ \citenamefont {Wang}(2017)}]{Huang2017}%
  \BibitemOpen
  \bibfield  {author} {\bibinfo {author} {\bibfnamefont {Li}~\bibnamefont
  {Huang}}\ and\ \bibinfo {author} {\bibfnamefont {Lei}\ \bibnamefont {Wang}},\
  }\bibfield  {title} {\enquote {\bibinfo {title} {Accelerated monte carlo
  simulations with restricted boltzmann machines},}\ }\href@noop {} {\bibfield
  {journal} {\bibinfo  {journal} {Phys. Rev. B}\ }\textbf {\bibinfo {volume}
  {95}},\ \bibinfo {pages} {035105} (\bibinfo {year} {2017})}\BibitemShut
  {NoStop}%
\bibitem [{\citenamefont {Liu}\ \emph {et~al.}(2017{\natexlab{a}})\citenamefont
  {Liu}, \citenamefont {Shen}, \citenamefont {Qi}, \citenamefont {Meng},\ and\
  \citenamefont {Fu}}]{Liu2016}%
  \BibitemOpen
  \bibfield  {author} {\bibinfo {author} {\bibfnamefont {Junwei}\ \bibnamefont
  {Liu}}, \bibinfo {author} {\bibfnamefont {Huitao}\ \bibnamefont {Shen}},
  \bibinfo {author} {\bibfnamefont {Yang}\ \bibnamefont {Qi}}, \bibinfo
  {author} {\bibfnamefont {Zi~Yang}\ \bibnamefont {Meng}}, \ and\ \bibinfo
  {author} {\bibfnamefont {Liang}\ \bibnamefont {Fu}},\ }\bibfield  {title}
  {\enquote {\bibinfo {title} {Self-learning monte carlo method and cumulative
  update in fermion systems},}\ }\href@noop {} {\bibfield  {journal} {\bibinfo
  {journal} {Phys. Rev. B}\ }\textbf {\bibinfo {volume} {95}},\ \bibinfo
  {pages} {241104} (\bibinfo {year} {2017}{\natexlab{a}})}\BibitemShut
  {NoStop}%
\bibitem [{\citenamefont {Liu}\ \emph {et~al.}(2017{\natexlab{b}})\citenamefont
  {Liu}, \citenamefont {Qi}, \citenamefont {Meng},\ and\ \citenamefont
  {Fu}}]{Liu2017}%
  \BibitemOpen
  \bibfield  {author} {\bibinfo {author} {\bibfnamefont {Junwei}\ \bibnamefont
  {Liu}}, \bibinfo {author} {\bibfnamefont {Yang}\ \bibnamefont {Qi}}, \bibinfo
  {author} {\bibfnamefont {Zi~Yang}\ \bibnamefont {Meng}}, \ and\ \bibinfo
  {author} {\bibfnamefont {Liang}\ \bibnamefont {Fu}},\ }\bibfield  {title}
  {\enquote {\bibinfo {title} {Self-learning monte carlo method},}\ }\href@noop
  {} {\bibfield  {journal} {\bibinfo  {journal} {Phys. Rev. B}\ }\textbf
  {\bibinfo {volume} {95}},\ \bibinfo {pages} {041101} (\bibinfo {year}
  {2017}{\natexlab{b}})}\BibitemShut {NoStop}%
\bibitem [{\citenamefont {Carleo}\ and\ \citenamefont
  {Troyer}(2017)}]{Carleo2017}%
  \BibitemOpen
  \bibfield  {author} {\bibinfo {author} {\bibfnamefont {Giuseppe}\
  \bibnamefont {Carleo}}\ and\ \bibinfo {author} {\bibfnamefont {Matthias}\
  \bibnamefont {Troyer}},\ }\bibfield  {title} {\enquote {\bibinfo {title}
  {Solving the quantum many-body problem with artificial neural networks},}\
  }\href@noop {} {\bibfield  {journal} {\bibinfo  {journal} {Science}\ }\textbf
  {\bibinfo {volume} {355}},\ \bibinfo {pages} {602} (\bibinfo {year}
  {2017})}\BibitemShut {NoStop}%
\bibitem [{\citenamefont {{Wang}}(2017)}]{Wang2017}%
  \BibitemOpen
  \bibfield  {author} {\bibinfo {author} {\bibfnamefont {L.}~\bibnamefont
  {{Wang}}},\ }\bibfield  {title} {\enquote {\bibinfo {title} {{Can Boltzmann
  Machines Discover Cluster Updates ?}}}\ }\href@noop {} {\bibfield  {journal}
  {\bibinfo  {journal} {ArXiv e-prints}\ } (\bibinfo {year} {2017})},\ \Eprint
  {http://arxiv.org/abs/1702.08586} {arXiv:1702.08586 [physics.comp-ph]}
  \BibitemShut {NoStop}%
\bibitem [{\citenamefont {{Gao}}\ and\ \citenamefont {{Duan}}(2017)}]{Gao2017}%
  \BibitemOpen
  \bibfield  {author} {\bibinfo {author} {\bibfnamefont {X.}~\bibnamefont
  {{Gao}}}\ and\ \bibinfo {author} {\bibfnamefont {L.-M.}\ \bibnamefont
  {{Duan}}},\ }\bibfield  {title} {\enquote {\bibinfo {title} {{Efficient
  Representation of Quantum Many-body States with Deep Neural Networks}},}\
  }\href@noop {} {\bibfield  {journal} {\bibinfo  {journal} {ArXiv e-prints}\ }
  (\bibinfo {year} {2017})},\ \Eprint {http://arxiv.org/abs/1701.05039}
  {arXiv:1701.05039 [cond-mat.dis-nn]} \BibitemShut {NoStop}%
\bibitem [{\citenamefont {{Deng}}\ \emph {et~al.}(2016)\citenamefont {{Deng}},
  \citenamefont {{Li}},\ and\ \citenamefont {{Das Sarma}}}]{Deng2016}%
  \BibitemOpen
  \bibfield  {author} {\bibinfo {author} {\bibfnamefont {D.-L.}\ \bibnamefont
  {{Deng}}}, \bibinfo {author} {\bibfnamefont {X.}~\bibnamefont {{Li}}}, \ and\
  \bibinfo {author} {\bibfnamefont {S.}~\bibnamefont {{Das Sarma}}},\
  }\bibfield  {title} {\enquote {\bibinfo {title} {{Exact Machine Learning
  Topological States}},}\ }\href@noop {} {\bibfield  {journal} {\bibinfo
  {journal} {ArXiv e-prints}\ } (\bibinfo {year} {2016})},\ \Eprint
  {http://arxiv.org/abs/1609.09060} {arXiv:1609.09060 [cond-mat.dis-nn]}
  \BibitemShut {NoStop}%
\bibitem [{\citenamefont {{Huang}}\ and\ \citenamefont
  {{Moore}}(2017)}]{Huang2017b}%
  \BibitemOpen
  \bibfield  {author} {\bibinfo {author} {\bibfnamefont {Y.}~\bibnamefont
  {{Huang}}}\ and\ \bibinfo {author} {\bibfnamefont {J.~E.}\ \bibnamefont
  {{Moore}}},\ }\bibfield  {title} {\enquote {\bibinfo {title} {{Neural network
  representation of tensor network and chiral states}},}\ }\href@noop {}
  {\bibfield  {journal} {\bibinfo  {journal} {ArXiv e-prints}\ } (\bibinfo
  {year} {2017})},\ \Eprint {http://arxiv.org/abs/1701.06246} {arXiv:1701.06246
  [cond-mat.dis-nn]} \BibitemShut {NoStop}%
\bibitem [{\citenamefont {{Levine}}\ \emph {et~al.}(2017)\citenamefont
  {{Levine}}, \citenamefont {{Yakira}}, \citenamefont {{Cohen}},\ and\
  \citenamefont {{Shashua}}}]{Levine2017}%
  \BibitemOpen
  \bibfield  {author} {\bibinfo {author} {\bibfnamefont {Y.}~\bibnamefont
  {{Levine}}}, \bibinfo {author} {\bibfnamefont {D.}~\bibnamefont {{Yakira}}},
  \bibinfo {author} {\bibfnamefont {N.}~\bibnamefont {{Cohen}}}, \ and\
  \bibinfo {author} {\bibfnamefont {A.}~\bibnamefont {{Shashua}}},\ }\bibfield
  {title} {\enquote {\bibinfo {title} {{Deep Learning and Quantum Entanglement:
  Fundamental Connections with Implications to Network Design}},}\ }\href@noop
  {} {\bibfield  {journal} {\bibinfo  {journal} {ArXiv e-prints}\ } (\bibinfo
  {year} {2017})},\ \Eprint {http://arxiv.org/abs/1704.01552} {arXiv:1704.01552
  [cs.LG]} \BibitemShut {NoStop}%
\bibitem [{\citenamefont {Kolmogorov}(1961)}]{Kolmogorov1961}%
  \BibitemOpen
  \bibfield  {author} {\bibinfo {author} {\bibfnamefont {A.~N.}\ \bibnamefont
  {Kolmogorov}},\ }\bibfield  {title} {\enquote {\bibinfo {title} {On the
  representation of continuous functions of several variables by superpositions
  of continuous functions of a smaller number of variables},}\ }\href@noop {}
  {\bibfield  {journal} {\bibinfo  {journal} {Doklady Akademii Nauk SSSR}\
  }\textbf {\bibinfo {volume} {108}},\ \bibinfo {pages} {179} (\bibinfo {year}
  {1961})}\BibitemShut {NoStop}%
\bibitem [{\citenamefont {Hornik}\ \emph {et~al.}(1989)\citenamefont {Hornik},
  \citenamefont {Stinchcombe},\ and\ \citenamefont {White}}]{Hornik1989}%
  \BibitemOpen
  \bibfield  {author} {\bibinfo {author} {\bibfnamefont {K.}~\bibnamefont
  {Hornik}}, \bibinfo {author} {\bibfnamefont {M.}~\bibnamefont {Stinchcombe}},
  \ and\ \bibinfo {author} {\bibfnamefont {H.}~\bibnamefont {White}},\
  }\bibfield  {title} {\enquote {\bibinfo {title} {Multilayer feedforward
  networks are universal approximators},}\ }\href@noop {} {\bibfield  {journal}
  {\bibinfo  {journal} {Neural networks}\ }\textbf {\bibinfo {volume} {2}},\
  \bibinfo {pages} {359} (\bibinfo {year} {1989})}\BibitemShut {NoStop}%
\bibitem [{\citenamefont {Cybenko}(1989)}]{Cybenko1989}%
  \BibitemOpen
  \bibfield  {author} {\bibinfo {author} {\bibfnamefont {G.}~\bibnamefont
  {Cybenko}},\ }\bibfield  {title} {\enquote {\bibinfo {title} {Approximation
  by superpositions of a sigmoidal function},}\ }\href@noop {} {\bibfield
  {journal} {\bibinfo  {journal} {Mathematics of control, signals and systems}\
  }\textbf {\bibinfo {volume} {2}},\ \bibinfo {pages} {303} (\bibinfo {year}
  {1989})}\BibitemShut {NoStop}%
\bibitem [{\citenamefont {Abadi}(2015)}]{Abadi2015}%
  \BibitemOpen
  \bibfield  {author} {\bibinfo {author} {\bibfnamefont {M.}~\bibnamefont
  {Abadi}},\ }\href@noop {} {\emph {\bibinfo {title} {Large-scale machine
  learning on heterogeneous systems}}}\ (\bibinfo  {publisher} {Software
  available from tensorflow.org},\ \bibinfo {year} {2015})\BibitemShut
  {NoStop}%
\bibitem [{\citenamefont {Saito}(2017)}]{Saito2017}%
  \BibitemOpen
  \bibfield  {author} {\bibinfo {author} {\bibfnamefont {Hiroki}\ \bibnamefont
  {Saito}},\ }\bibfield  {title} {\enquote {\bibinfo {title} {Solving the
  bose-hubbard model with machine learning},}\ }\href@noop {} {\bibfield
  {journal} {\bibinfo  {journal} {J. Phys. Soc. Jpn.}\ }\textbf {\bibinfo
  {volume} {86}},\ \bibinfo {pages} {093001} (\bibinfo {year}
  {2017})}\BibitemShut {NoStop}%
\bibitem [{\citenamefont {Saito}\ and\ \citenamefont {Kato}(2018)}]{Saito2018}%
  \BibitemOpen
  \bibfield  {author} {\bibinfo {author} {\bibfnamefont {Hiroki}\ \bibnamefont
  {Saito}}\ and\ \bibinfo {author} {\bibfnamefont {Masaya}\ \bibnamefont
  {Kato}},\ }\bibfield  {title} {\enquote {\bibinfo {title} {Machine learning
  technique to find quantum many-body ground states of bosons on a lattice},}\
  }\href@noop {} {\bibfield  {journal} {\bibinfo  {journal} {J. Phys. Soc.
  Jpn.}\ }\textbf {\bibinfo {volume} {87}},\ \bibinfo {pages} {014001}
  (\bibinfo {year} {2018})}\BibitemShut {NoStop}%
\bibitem [{\citenamefont {{Kingma}}\ and\ \citenamefont
  {{Ba}}(2014)}]{Kingma2014}%
  \BibitemOpen
  \bibfield  {author} {\bibinfo {author} {\bibfnamefont {D.~P.}\ \bibnamefont
  {{Kingma}}}\ and\ \bibinfo {author} {\bibfnamefont {J.}~\bibnamefont
  {{Ba}}},\ }\bibfield  {title} {\enquote {\bibinfo {title} {{Adam: A Method
  for Stochastic Optimization}},}\ }\href@noop {} {\bibfield  {journal}
  {\bibinfo  {journal} {ArXiv e-prints}\ } (\bibinfo {year} {2014})},\ \Eprint
  {http://arxiv.org/abs/1412.6980} {arXiv:1412.6980 [cs.LG]} \BibitemShut
  {NoStop}%
\end{thebibliography}
%


\end{document}